\documentclass[paper]{myJHEP3}
\usepackage{amsmath}
\usepackage{comment}
\usepackage{slashed}
\usepackage{graphicx}
\usepackage[english]{babel}
\usepackage{cite}

\newcommand{\M}{\,\mathcal{M}}
\newcommand{\F}{\,\mathcal{F}}
\newcommand{\iA}[2]{ \,\big[ \,#1 \,m \,|\, #2 \,m \,\big] }
\newcommand{\iB}[2]{ \,\big[ \,#1 \,m\,|\, #2 \,m\,|\, 0 \,\big] }
\newcommand{\fdru}{[\de^4q_1]}
\newcommand{\fdrd}{[\de^4q_2]}
\newcommand{\fdr}{[\de^4q]}
\newcommand{\de}{\mathrm{d}}
\newcommand{\eps}{\varepsilon}
\newcommand{\beq}{\begin{equation}}
\newcommand{\eeq}{\end{equation}}
\newcommand{\bqa}{\begin{eqnarray}}
\newcommand{\eqa}{\end{eqnarray}}

\newcommand{\nl}{\nonumber \\}
\newcommand{\iO}[1]{ \,\big[ \,#1 \,m\,\big] }

\newcommand{\dbar}{\bar D}
\newcommand{\sbar}{\bar S}

\newcommand{\qbar}{\bar q}

\def\spa#1.#2{\left\langle#1\,#2\right\rangle}
\def\spb#1.#2{\left[#1\,#2\right]}

\def\feynsl#1{
  \setbox0=\hbox{/} \setbox1=\hbox{$#1$}
  \dimen0=\wd0 \advance\dimen0 by -\wd1 \divide\dimen0 by 2
  \ifdim\wd0>\wd1 \raise.15ex\copy0\kern-\wd0\kern\dimen0\copy1\kern\dimen0
  \else \kern-\dimen0\raise.15ex\copy0\kern-\dimen0\kern-\wd1\copy1\fi}

\def\ord{{\cal O} }

\newcommand\sss{\scriptscriptstyle}

\newcommand\mur{\mu_{\sss\rm R}}

\newskip\humongous \humongous=0pt plus 1000pt minus 100pt

\newif\ifdtup

\def    \br(#1,#2)          {\mbox{$\langle #1 \, #2 \rangle$}}
\def    \sq(#1,#2)          {\mbox{$\left[  #1 \, #2 \right]$}}

\title{FDR, an easier way to NNLO calculations: a two-loop case study}

\author{Alice Maria Donati and Roberto Pittau\\
        Departamento de F\'isica Te\'orica y del Cosmos and CAFPE,
  Campus Fuentenueva s. n., Universidad de Granada, E-18071 Granada, Spain \\
  E-mail: \email{adonati@ugr.es,pittau@ugr.es}}

\abstract{In this paper we illustrate the simplifications produced by FDR in NNLO computations. We show with an explicit example that --\,due to its four-dimensionality\,-- FDR does not require an order-by-order renormalization and that, unlike the one-loop case, FDR and dimensional regularization (DR) generate intermediate two-loop results which are no longer linked by a simple subtraction of the ultraviolet (UV) poles in $\epsilon$. 
Our case study is the two-loop amplitude for $H \to \gamma \gamma$, mediated by an infinitely heavy top loop, in the presence of gluonic corrections. We use this to elucidate how gauge invariance is preserved with no need of introducing counterterms in the Lagrangian.
In addition, we discuss a possible four-dimensional approach to the infrared (IR) problem compatible with the FDR treatment of the UV infinities.}

\preprint{}

\begin{document}
\section{Introduction}
Computing radiative corrections has become of uppermost importance in particle phenomenology~\cite{Blumlein:2012gu}.
The present lack of unexpected signals at the LHC pulls the effects of New Physics in a domain where small discrepancies have to be searched via detailed comparisons between experimental results and precise calculations of the Standard Model background. Due to the large QCD coupling constant, precise predictions at the LHC often require NNLO accuracy. 
On the other hand, two-(or more)-loop calculations in the complete Electroweak (EW) model will be mandatory at the future International Linear Collider to meet the experimental accuracy foreseen, for example, in Higgs Physics~\cite{Simon:2014aqa}.

While NLO techniques are very well established~\cite{Passarino:1978jh,Frixione:1995ms,Catani:1996vz,Kosower:1997zr,Campbell:1998nn,Catani:2002hc,Nagy:2003qn,Bern:1994cg,Britto:2004nc,Ossola:2006us,Forde:2007mi,Berger:2008sj,Ellis:2009zw}, work is ongoing to solve the NNLO problem in its full generality~\cite{Passarino:2001wv}. As for the virtual sector, progress has been recently achieved by extending generalized unitarity techniques at two-loops~\cite{Mastrolia:2011pr,Mastrolia:2012an,Kleiss:2012yv,Badger:2012dp,Johansson:2012zv}, while the antenna subtraction~\cite{GehrmannDeRidder:2005cm,Currie:2013dwa} and sector decomposition~\cite{Binoth:2000ps,Anastasiou:2003gr,Binoth:2004jv,Czakon:2010td} methods look promising tools to deal with IR divergences beyond NLO~\cite{Catani:2007vq}.

In this paper we investigate the possibility of further simplifying NNLO computations by abandoning dimensional regularization~\cite{'tHooft:1972fi}. Despite its known virtues, DR requires a heavy analytic work aimed at subtracting powers of $1/\epsilon$ of UV or IR origin even before attacking the calculation of the finite physical part. For instance, DR forces an order-by-order iterative renormalization, which is especially cumbersome when computing loop corrections in the EW model or in SUSY: the full set of one-loop counterterms has to be determined and added in a two-loop computation, and so on. Furthermore, loop functions used at a given perturbative level must be further expanded in $\epsilon$ --\,when appearing at higher orders\,-- to include terms generating ${\cal O}(\epsilon^0)$ contributions when multiplied by the new poles.  Such complications arise in DR because constants needed to preserve the symmetries of the Lagrangian are often produced by $\epsilon/\epsilon$ terms, which are kept under control by the iterative renormalization. 

This has driven us to study the performances of FDR~\cite{Pittau:2012zd} as a simpler four-dimensional approach beyond one loop\,\footnote{Other four-dimensional treatments are listed in~\cite{Freedman:1991tk,delAguila:1997kw,delAguila:1998nd,Battistel:1998sz,Cynolter:2010ei,Cherchiglia:2010yd}.}.
The key point of FDR is that the use of counterterms is avoided by {\it defining} a four-dimensional and UV-free loop integration in a way compatible with shift and gauge invariance. Having done this, the correct results automatically emerge once the theory is fixed in terms of physical observables by means of a {\it finite} renormalization relating the parameters of the Lagrangian to measured quantities. In addition, IR infinities can be naturally accommodated within the same four-dimensional framework used to cope with the UV divergences.
This is why we envisage in FDR a great potential to reduce the complexity of the NNLO calculations, especially when used together with numerical techniques.
 
In this paper we present, as the first example of a two-loop FDR calculation, the QCD corrections to the top-loop-mediated Higgs decay into two photons, in the limit $m_{\rm top} \to \infty$. This computation gives the opportunity to fully appreciate the simplifications due to the four-dimensionality of the approach in a realistic two-loop case study\,\footnote{One-loop examples have been worked out in~\cite{Pittau:2013qla,Donati:2013iya}.}.
In the next section we review the general FDR idea with special emphasis on the two-loop case. We discuss, in particular, the shift and gauge invariance properties of the FDR integration, the main differences with DR, and the IR problem. The two-loop FDR calculation of $H \to \gamma \gamma$ is presented in section~\ref{calculation} and the technical details are collected in the final appendices.

\section{FDR and the importance of working in four dimensions} 
\subsection{Definition of the FDR loop integral}
FDR subtracts UV divergences at the integrand level. This is obtained in two steps.  Firstly, the propagators of the particles flowing in the loops are given a common additional term $-\mu^2$, formally identified with the $+ i 0$ propagator prescription. For example, vector-boson and fermion propagators with momentum $(q+p_i)$ and mass $M_i$ read\,\footnote{$q$ denotes a generic integration momentum and $p_i$ and external momentum.}, in the unitary gauge, 
\bqa
\label{eqpr}
\frac{g^{\alpha \beta}-(q+p_i)^\alpha(q+p_i)^\beta/M^2_i}{\dbar_{p_i}}~~~~{\rm and}~~~~
\frac{\rlap /q + \rlap /p_i + M_i}{\dbar_{p_i}}\,,
\eqa
respectively, with
\bqa
\dbar_{p_i} &=& (q+p_i)^2-M^2_i-\mu^2~=~ \qbar^2-d_i\,,\nl
\qbar^2 &\equiv& ~q^2-\mu^2\,,~~
d_i ~\equiv~ M^2_i-p^2_i- 2 (q\cdot p_i)\,.
\eqa
Secondly, UV infinities are isolated by a repeated use of the identity
\bqa
\label{eq:id}
	\frac{1}{\dbar_{p_i}}  = \frac{1}{\qbar^2}
		     +\frac{d_i}{\qbar^2 \dbar_{p_i}}\,.
\eqa
In fact --\,being $d_i$ is at most linear in $q$\,-- the second term in the r.h.s. of Eq.~(\ref{eq:id}) is less UV divergent than the original denominator, so that UV divergences can be systematically moved to terms such as ${1}/{\bar q^2}$, which {\em depend only on $\mu$}, and directly subtracted from the integrand. Schematically, dubbing $J$ the original integrand of an $\ell$-loop function, one has
\bqa
\label{eq:fdrexpansion}
J(q_1,\ldots, q_\ell)= J_{\rm INF}(q_1,\ldots, q_\ell)+J_{{\rm F},\ell}(q_1,\ldots, q_\ell)\,,
\eqa
where $J_{\rm INF}$ collects the UV divergent integrands. Then, the FDR integral over $J$ is {\it defined} as\,\footnote{Throughout the paper FDR integration is denoted by the symbol $[d^4q_i]$.}
\bqa
\label{eq:fdrdef}
\int [d^4q_1] \ldots [d^4q_\ell]\,  J(q_1,\ldots, q_\ell) \equiv \lim_{\mu \to 0}
\int d^4q_1 \ldots d^4q_\ell\, J_{{\rm F},\ell}(q_1,\ldots q_\ell)\,,
\eqa
where, due to the limit ${\mu \to 0}$, only a logarithmic dependence on $\mu$ remains, which can be traded for a dependence on the renormalization 
scale\,\footnote{See subsection~\ref{indcut}.}.
Thus, FDR and normal integration coincide in a convergent integral, since no divergent part $J_{\rm INF}$ can be extracted from its integrand. 
Furthermore, the space-time is kept strictly four-dimensional also in divergent integrals --\,with $g^{\alpha \beta}= {\rm diag}(1,-1,-1,-1)$\,-- because $\mu^2$ is nothing but the infinitesimal deformation needed to define the loop integrals\,\footnote{Unlike in DR, the limit $\mu \to 0$ is taken outside integration (see Eq.~(\ref{eq:fdrdef})).} and it is not generated by higher-dimensional components of the integration momenta. This allows one to perform, in particular, the Dirac gamma algebra in four dimensions, with extra rules needed to keep gauge invariance, as explained in subsection~\ref{subsec:gauge}.

An explicit example of integrand FDR expansion\,\footnote{We denote the expansion of an integrand $J$ needed to bring it
in the form of Eq.~(\ref{eq:fdrexpansion}) as its {\em FDR defining expansion}.}
at one loop is given by
\bqa
\label{eq:ex1}
\!{\frac{q^\alpha q^\beta}{\dbar_{p_0} \dbar_{p_1}}}   &=&    
 {  \left[\frac{q^\alpha q^\beta}{\qbar^4} \right]}\!
+ (d_0+M^2_1-p^2_1)\!\left[\frac{q^\alpha q^\beta}{\qbar^6} \right]\!
-2 p_{1\gamma}\! \left[\frac{q^\alpha q^\beta q^\gamma}{\qbar^6} \right]\!
\nl &&
+4 p_{1\gamma} p_{1\delta} \left[\frac{q^\alpha q^\beta q^\gamma q^\delta}{\qbar^8} \right]\!
+ J^{\alpha \beta}_{{\rm F},1}(q)\,,\nl
J^{\alpha \beta}_{{\rm F},1}(q)&=& q^\alpha q^\beta \left(
\frac{4 (q \cdot p_1)^2 d_1}{\qbar^8 \dbar_{p_1}}
+ (M_1^2-p_1^2) \frac{d_0+d_1-2(q \cdot p_1)}{\qbar^6 \dbar_{p_1}}
\right. \nl
&& -2 d_0 \frac{(q \cdot p_1)}{\qbar^6 \dbar_{p_1}}  + \left. \frac{d_0^2}{\qbar^4 \dbar_{p_0}\dbar_{p_1}}
\right)\,, 
\eqa
where $p_0= 0$ and the terms in square brackets are proportional to UV divergent integrands. 
A two-loop example with  
\bqa
\label{eq:2lden}
\bar D_1   = \bar{q}_1^2-m_1^2\,,~~
\bar D_2   = \bar{q}_2^2-m_2^2\,,~~
\bar D_{12} = \bar{q}_{12}^2-m_{12}^2\,,~~
q_{12}= q_1+q_2
\eqa
reads
\bqa
\label{eq:ex2}
\frac{1}{\bar D_1\bar D_2\bar D_{12}} &=&
{ \left[\frac{1}{\bar{q}_1^2\bar{q}_2^2 \bar{q}_{12}^2}\right]} 
+ m_1^2
{ \left[
\frac{1}{\bar{q}_1^4\bar{q}_2^2 \bar{q}_{12}^2}
 \right]}
+ m_2^2
{ \left[
\frac{1}{\bar{q}_1^2 \bar{q}_2^4\bar{q}_{12}^2}
 \right]}
+
 m_{12}^2
{
\left[
\frac{1}{\bar{q}_1^2 \bar{q}_2^2\bar{q}_{12}^4}
 \right]}
\nl
&+&
\frac{m_1^4}{(\bar D_1\bar{q}_1^4)}
{ \left[\frac{1}{\bar{q}_2^4} \right]}
+ 
\frac{m_2^4}{(\bar D_2\bar{q}_2^4)}
{ \left[\frac{1}{\bar{q}_1^4} \right]}
+
 \frac{m_{12}^4}{(\bar D_{12}\bar{q}_{12}^4)}
{ \left[\frac{1}{\bar{q}_1^4} \right]} 
\nl&+&
J_{{\rm F},2}(q_1,q_2) \,,
\eqa
where
\bqa
J_{{\rm F},2}(q_1,q_2) &=&
- m_1^4 \frac{q_1^2+2(q_1 \cdot q_2)}{(\bar D_1 \bar{q}_1^4)\bar{q}_2^4 \bar{q}_{12}^2}
- m_2^4 \frac{q_2^2+2(q_1 \cdot q_2)}{\bar{q}_1^4 (\bar D_2 \bar{q}_2^4) \bar{q}_{12}^2}
- m_{12}^4 \frac{q_{12}^2-2(q_1 \cdot q_{12})}{\bar{q}_1^4 \bar{q}_{2}^2
(\bar D_{12} \bar{q}_{12}^4)}
\nonumber \\
&+& 
 \frac{m_1^2 m_2^2}{(\bar D_1 \bar{q}_1^2)(\bar D_2 \bar{q}_2^2)\bar{q}_{12}^2} 
+\frac{m_1^2 m_{12}^2}{(\bar D_1 \bar{q}_1^2)\bar{q}_{2}^2(\bar D_{12} \bar{q}_{12}^2)} 
+\frac{m_2^2 m_{12}^2}{\bar{q}_{1}^2(\bar D_2 \bar{q}_2^2)(\bar D_{12} \bar{q}_{12}^2)} 
\nonumber \\
&+& 
\frac{m_1^2 m_2^2 m_{12}^2}{(\bar D_1 \bar{q}_1^2)(\bar D_2 \bar{q}_2^2)(\bar D_{12} \bar{q}_{12}^2)}\,.
\eqa
Note that identities such as
\bqa
\frac{1}{\qbar^2_{12}}= \frac{1}{\qbar^2_2}-\frac{q^2_1+2 (q_1 \cdot q_2)}{\qbar^2_2 \qbar^2_{12} }\,,~~~~~
\frac{1}{\qbar^2_{2}}= \frac{1}{\qbar^2_1}-\frac{q^2_{12}-2 (q_1 \cdot q_{12})}{\qbar^2_1 \qbar^2_2 }
\eqa
are needed to extract the sub-divergences.
Then, the  one- and two-loop FDR integrals over the integrands in
Eqs.~(\ref{eq:ex1}) and (\ref{eq:ex2})  read
\bqa
\int [d^4q]  \frac{q^\alpha q^\beta}{\bar D_{p_0}\bar D_{p_1}} &=& \lim_{\mu \to 0}
\int d^4q\, J^{\alpha \beta}_{{\rm F},1}(q)\,,  \nl
\int [d^4q_1] [d^4q_2]  \frac{1}{\bar D_1\bar D_2\bar D_{12}}
&=& \lim_{\mu \to 0} \int d^4q_1 d^4q_2\,J_{{\rm F},2}(q_1,q_2)\,.  
\eqa

It is important to realize that divergent tensor structures are fully subtracted from the original integrand, as in Eq.~(\ref{eq:ex1})\,\footnote{It can be shown that FDR tensors are equivalent to DR tensors at one loop, but differences start at two loops and beyond~\cite{Pittau:2013ica}.}.
Owing to the Lorentz invariance and four-dimensionality of this definition, irreducible tensors can be decomposed in terms of scalars. For example\,\footnote{The FDR defining expansion of  
$\frac{q^\alpha_1 q^\beta_1}{\bar D_1^3\bar D_2\bar D_{12}}$ is given in appendix~\ref{appa}.}
\bqa
\label{eq:extens}
&&\int [d^4q_1] [d^4q_2] \frac{q^\alpha_1 q^\beta_1}{\bar D_1^3\bar D_2\bar D_{12}}
 =  
\lim_{\mu \to 0} 
\int d^4q_1 d^4q_2\,\,q^\alpha_1 q^\beta_1\,J_D(q_1,q_2)\,,
~~~{\rm where} \nl 
&&J_D(q_1,q_2)  =  \bigg\{
  \left(\frac{1}{\bar{q}_1^6}
       - \frac{1}{\bar D^3_1}\right)
  \frac{q_1^2+2(q_1 \cdot q_2)}{\bar{q}_2^4 \bar{q}_{12}^2}
  \nl && \qquad\qquad\qquad\qquad\qquad\qquad
 +\frac{1}{\bar D_{1}^3 \bar{q}_2^2\bar D_{12} }
 \left(
 \frac{m_2^2}{\bar D_{2}}+\frac{m_{12}^2}{\bar{q}_{12}^2}
 \right) 
  \bigg\}\,,
\eqa
can be rewritten as
\bqa
\label{eq:simpl1}
\int [d^4q_1] [d^4q_2] \frac{q^\alpha_1 q^\beta_1}{\bar D_1^3\bar D_2\bar D_{12}}
=
\frac{g^{\alpha \beta }}{4}\int [d^4q_1] [d^4q_2] \frac{q^2_1}{\bar D_1^3\bar D_2\bar D_{12}}\,,
\eqa
with
\bqa
\label{eq:simpl2}
\int [d^4q_1] [d^4q_2] \frac{q^2_1}{\bar D_1^3\bar D_2\bar D_{12}}
= \lim_{\mu \to 0} 
\int d^4q_1 d^4q_2\,\,q^2_1\,J_D(q_1,q_2)\,.
\eqa

Finally, polynomials in the integration variables represent a limiting case of Eq.~(\ref{eq:fdrexpansion}), in which 
\bqa
J_{{\rm F},\ell}(q_1,\ldots, q_\ell)= 0\,. 
\eqa
As a consequence
\bqa
\int [d^4q] \left(\qbar^2\right)^\alpha = 0\,,
\eqa
for any integer $\alpha \geq 0$.

\subsection{Shift invariance and uniqueness}
FDR integrals are invariant under the shift of any integration variable. 
This can be easily proven by using the fact that they can be thought as finite differences of shift-invariant dimensionally-regulated\,\footnote{Here and in the following $n= 4 + \epsilon$ and $\mur$ is the renormalization scale.} divergent integrals (see Eq. (\ref{eq:fdrexpansion})) 
\bqa
\label{eqdiffn}
\int [d^4q_1] \ldots [d^4q_\ell]\,  J(q_1,\ldots, q_\ell) &=& 
\lim_{\mu \to 0} \mur^{-\ell \epsilon}
\left(
 \int d^nq_1 \ldots d^nq_\ell \,  J(q_1,\ldots, q_\ell) \right. \nl
&& \left. -\int d^nq_1 \ldots d^nq_\ell \,  J_{\rm INF}(q_1,\ldots, q_\ell)
\right)\,.
\eqa 
The explicit demonstration is given in appendix~\ref{appshift}. A corollary to this theorem is the uniqueness of the definition in Eq.~(\ref{eq:fdrdef}). In fact, the subtracted integrands in $J_{\rm INF}(q_1,\ldots, q_\ell)$ are unambiguously determined by the UV content of the original integrand, the only possible ambiguity being shifts of the loop momenta in $J(q_1,\ldots, q_\ell)$, which, however, produce the same FDR integral.

Eq.~(\ref{eqdiffn}) also demonstrates that whenever DR loop integrals are known, their FDR counterparts can also be computed.   

\subsection{Independence of the cutoff}
\label{indcut}
As a result of the subtraction of the divergent integrands, non integrable powers of $1/\qbar^2$ are developed in $J_{{\rm F},\ell}(q_1,\ldots, q_\ell)$. Such IR poles get regulated by the $\mu^2$ propagator prescription, which gives a meaning to the the r.h.s. of Eq.~(\ref{eq:fdrdef}). Thus, the original UV cutoff is traded for an IR one: $\mu$. Here we show that FDR integrals depend at most logarithmically on $\mu$. Furthermore, $\mu$ can be traded for the renormalization scale $\mur$, rendering the definition of the FDR integration independent of any cutoff.  

We start from Eq.~(\ref{eqdiffn}). Since the first term in its r.h.s. is the original DR regulated integral it does not depend on $\mu$, in the limit $\mu \to 0$\,\footnote{This is true in the absence of IR divergences. However, UV and IR infinities simultaneously occur only in scale-less integrals, which vanish in FDR (see subsection~\ref{ircl}).}.  On the other hand, polynomially divergent integrands in $J_{\rm INF}$ cannot contribute either, because they generate polynomials in $\mu$. Therefore, the $\mu$ dependence in the l.h.s. is entirely due to powers of $\ln(\mu/\mur)$ created by the subtraction of the logarithmically divergent integrals. If one redefines FDR integrals without subtracting such logarithms, no dependence on $\mu$ is produced.
This is equivalent to the operation of adding back all $\ln(\mu/\mur)$s to the l.h.s. of Eq.~(\ref{eqdiffn}). Then, the limit ${\mu \to 0}$ can be taken, $\mu$ becomes $\mur$ and no cutoff is left. The identification $\mu = \mur$ after 
$\lim_{\mu \to 0}$ is understood in all FDR integrals appearing in this paper. 
 
\subsection{Keeping gauge invariance}
\label{subsec:gauge}
Now we discuss how gauge invariance is preserved in FDR. Our starting point is the existence of graphical proofs of the Ward-Slavnov-Taylor identities~\cite{Sterman:1994ce}, in which the correct relations among Green's functions are demonstrated --\,at any loop order\,-- directly at the level of Feynman diagrams.
Such proofs are valid under two circumstances:
\begin{itemize}
\item divergent loop integrals should be defined in a way that shifting the integration momenta is possible as if they were convergent ones~\cite{Collins:1984xc};
\item cancellations between numerators and denominators should be 
preserved\,\footnote{Quoting Martinus Veltman~\cite{veltman74}:
{\em Gauge invariance implies a tight interplay between the numerator of an 
integrand and its denominator. Changing either of the two will generally destroy gage invariance $\ldots$}}. 
\end{itemize}
Since the first property has been already proven, we concentrate here on the second requirement, which we study by means of a two-loop example.

Consider the scalar integral
\bqa
\int [d^4q_1] [d^4q_2] 
  \frac{1}{\bar D_1^2\bar D_2\bar D_{12}}\,.
\eqa
To define it in FDR, it is necessary to make explicit the $\mu^2$ dependence in its denominators\,\footnote{See Eq.~(\ref{eq:2lden}).}, which amounts to the replacement 
\bqa
\label{eq:deform}
q^2_i \to q^2_i -\mu^2.
\eqa
However, this change should be performed without altering the cancellations which ensure that the same result is obtained   both by
simplifying the reducible numerators before computing the integrals and by  working out the integrals without simplifying the numerators.
That happens only if
\begin{enumerate}
\item 
any $q^2_i$ generated by Feynman rules in the numerator of a diagram\,\footnote{Such $q^2$ terms are created, for instance, when $(q+p_i)^\alpha(q+p_i)^\beta/M^2_i$ and $\rlap /q + \rlap /p_1 + M_i$
in Eq.~(\ref{eqpr}) are multiplied by $g_{\alpha \beta}$ and $\rlap/q$ respectively, before tensor reduction.} is also changed as in Eq.~(\ref{eq:deform});
\item 
simplifications at the integrand level are possible, such as
\bqa
\label{2loopFDR}
\int [d^4q_1] [d^4q_2] 
  \frac{q^2_1 -m^2_1-\mu^2}{\bar D_1^3\bar D_2\bar D_{12}} &=&
\int [d^4q_1] [d^4q_2] 
  \frac{1}{\bar D_1^2\bar D_2\bar D_{12}}\,.
\eqa
\end{enumerate}
Either way, integrals with $\mu$ in the numerator appear --\,which we dub {\it extra integrals}\,-- that need to be properly defined. For instance, since an explicit computation gives
\bqa
\label{2loopFDRpp}
\int [d^4q_1] [d^4q_2] 
  \frac{q^2_1-m^2_1}{\bar D_1^3\bar D_2\bar D_{12}} &\ne&
\int [d^4q_1] [d^4q_2] 
  \frac{1}{\bar D_1^2\bar D_2\bar D_{12}}\,,
\eqa
one deduces that\,\footnote{The r.h.s. of Eq.~(\ref{2loopFDRppp}) vanishes because FDR integrals are at most logarithmic in $\mu$.}
\bqa
\label{2loopFDRppp}
\int [d^4q_1] [d^4q_2] 
  \frac{\mu^2}{\bar D_1^3\bar D_2\bar D_{12}} \ne
\lim _{\mu \to 0}\mu^2 \int [d^4q_1] [d^4q_2] 
  \frac{1}{\bar D_1^3\bar D_2\bar D_{12}} = 0\,.
\eqa
In fact, a non-zero contribution must be added to the l.h.s. of Eq.~(\ref{2loopFDRpp}) to produce the r.h.s. of Eq.~(\ref{2loopFDR}).
The right cancellation occurs if the denominators $1/{\bar D_1^3\bar D_2\bar D_{12}}$ are expanded in front of $\mu^2$ {\it as if it was a $q^2_1$}, namely as in Eq.~(\ref{eq:extens})\,\footnote{\label{foot}It is interesting to study how a finite contribution is generated by the definition in Eq.~(\ref{eq:eqxtra1}). In $J_D(q_1,q_2)$
\bqa
\label{eq:1mu2}
\int d^4q_1 d^4q_2\,
  \frac{q_1^2+2(q_1 \cdot q_2)}{\qbar_1^6 \bar{q}_2^4 \bar{q}_{12}^2} \sim \frac{1}{\mu^2}\,, \nonumber
\eqa
thus
\bqa
\label{eq:rightconst1}
\int [d^4q_1] [d^4q_2] \frac{\mu^2|_1}{\bar D_1^3\bar D_2\bar D_{12}}
=  \lim_{\mu \to 0} \mu^2\!\! \int d^4q_1 d^4q_2
  \frac{q_1^2+2(q_1 \cdot q_2)}{\qbar_1^6 \bar{q}_2^4 \bar{q}_{12}^2}\, \nonumber
\eqa
produces a finite constant when $\mu \to 0$.
The value of this integral is given in subsection~\ref{sec:bblo}.}:
\bqa
\label{eq:eqxtra1}
\int [d^4q_1] [d^4q_2] 
  \frac{\mu^2|_1}{\bar D_1^3\bar D_2\bar D_{12}} =  
\lim_{\mu \to 0} 
\int d^4q_1 d^4q_2\,\,\mu^2\,J_D(q_1,q_2)\,.
\eqa
By using this definition, Eq.~(\ref{2loopFDR}) directly follows from the FDR defining expansion of its two sides.
Note that the index $_1$ in $\mu^2|_1$ only denotes the expansion to be used:
although only one kind of $\mu^2$ exists
\bqa
\label{eq:eqextra}
\int [d^4q_1] [d^4q_2] \frac{\mu^2|_1}{\bar D_1^3\bar D_2\bar D_{12}}\,,~
\int [d^4q_1] [d^4q_2] \frac{\mu^2|_2}{\bar D_1^3\bar D_2\bar D_{12}}\,
\nl~{\rm and}~
\int [d^4q_1] [d^4q_2] \frac{\mu^2|_{12}}{\bar D_1^3\bar D_2\bar D_{12}}
\eqa 
are in general different, because they are defined by expanding 
\bqa
\int [d^4q_1] [d^4q_2] \frac{q^2_1}{\bar D_1^3\bar D_2\bar D_{12}}\,,~
\int [d^4q_1] [d^4q_2] \frac{q^2_2}{\bar D_1^3\bar D_2\bar D_{12}}\,
\nl~{\rm and}~
\int [d^4q_1] [d^4q_2] \frac{q^2_{12}}{\bar D_1^3\bar D_2\bar D_{12}}\,,
\eqa
respectively.

The described procedure is completely general: the extra integrals are defined by the FDR expansion of the integrals obtained by replacing $\mu|_i \to q_i$. 
As a consequence, the $\mu|_i$ in the numerator are sensitive to changes of variables. For example, if $q_1 \to q_1 -q_2$ and $q_2 \to -q_2$,
\bqa
\int [d^4q_1] [d^4q_2] \frac{\mu^2|_{12}}{\bar D_1^3\bar D_2\bar D_{12}}~~\to~~
\int [d^4q_1] [d^4q_2] \frac{\mu^2|_1}{\bar D_1 \bar D_2\bar D_{12}^3}\,.
\eqa

Extra integrals can be computed either {\em directly}, by considering 
the finite part $J_D(q_1,q_2)$ of the relevant denominator expansion 
--\,as done in Eq.~(\ref{eq:eqxtra1})\,-- or {\em indirectly}, by rewriting 
$J_D(q_1,q_2)$ as a difference between the original integrand and its subtracted pieces. This second way is usually more convenient, because 
the original integral does not contribute in the limit $\mu \to 0$. 
For example, Eq.~(\ref{eq:appa1}) gives
\bqa
\int [d^4q_1] [d^4q_2] \frac{\mu^2|_1}{\bar D_1^3\bar D_2\bar D_{12}} =
-\lim_{\mu \to 0} \mu^2
\int d^4q_1 d^4q_2
\left(
\frac{1}{\qbar^6_1 \qbar^2_2 \qbar^2_{12}}  
  -
       \frac{1}{\bar{q}_1^6}
\frac{1}{\qbar_2^4}
\right) \,,
\eqa
which coincides with the result in footnote~\ref{foot}.

Finally, extra integrals give the possibility to rewrite tensors in terms of scalars plus constants. For instance, Eq.~(\ref{eq:simpl1}) produces
\bqa
\label{eq:tensdec}
\int [d^4q_1] [d^4q_2] 
  \frac{q^\alpha_1q^\beta_1}{\bar D_1^3\bar D_2\bar D_{12}} &=&
\frac{g^{\alpha \beta}}{4} \left\{
\int [d^4q_1] [d^4q_2] 
 \frac{1}{\bar D_1^2\bar D_2\bar D_{12}} 
\right. \nl && + \left.
m_1^2
\int [d^4q_1] [d^4q_2] 
  \frac{1}{\bar D_1^3\bar D_2\bar D_{12}} 
\right. \nl && + \left. 
\int [d^4q_1] [d^4q_2] 
  \frac{\mu^2|_1}{\bar D_1^3\bar D_2\bar D_{12}}
\right\}\,. 
\eqa
Decompositions like this will be extensively used in the calculation presented in section~\ref{calculation}.

\begin{figure}[t]
\begin{center}
\includegraphics[width=0.8\textwidth]{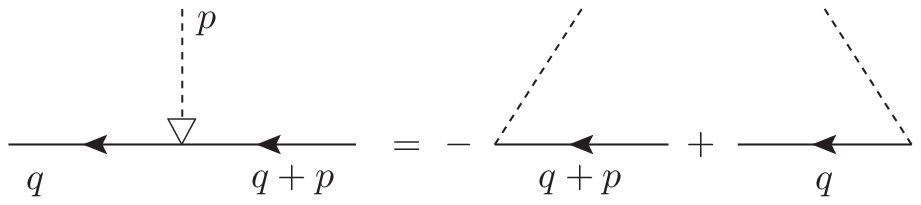}
\caption{\label{qedwi} Graphical representation of the Feynman identity 
in Eq.~(\ref{feyd}). The dashed line represents a scalar photon.}
\end{center}
\end{figure}

\begin{figure}[t]
\begin{center}
\includegraphics[width=0.6\textwidth]{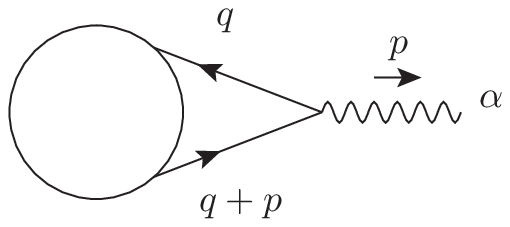}
\caption{\label{genamp} Generic $\ell$-loop amplitude with an external photon with momentum $p$. The blob stands for the rest of the amplitude and $q$ is an integration momentum.}
\end{center}
\end{figure}

Having studied the general mechanism of the gauge cancellations in FDR, we further elucidate it by means of the process investigated in this paper, namely $H \to \gamma \gamma$ mediated by a fermion with mass $m$. In this case the proof of gauge invariance relies on the graphical equivalence depicted in Fig.~\ref{qedwi}, 
which, in turn, is realized by the Feynman identity
\bqa
\label{feyd}
\frac{\rlap/q +m}{D}\, \rlap/p\, \frac{\rlap/q+\rlap/p +m}{D_p}=
\frac{\rlap/q +m}{D}       - \frac{\rlap/q+\rlap/p +m}{D_p}\,,
\eqa
where
\bqa
D = q^2-m^2~~~{\rm and}~~~D_p= (q+p)^2-m^2\,.
\eqa
Consider now the generic $\ell$-loop amplitude in Fig.~\ref{genamp}.
Its integrand reads 
\bqa
\label{eq:integg}
\epsilon_\alpha(p) J^\alpha(q,\cdots,q_{\ell}) &=&
\epsilon_\alpha(p) 
\sum_i \frac{1}{D D_p}
\nl && \times
Tr \left[(\rlap /q +m) \gamma^\alpha (\rlap /q+\rlap /p +m)
\left(\Gamma^{i}_o + \Gamma^{i}_e 
\right)   
 \right]\,,
\eqa
where the sum is over all contributing Feynman diagrams and 
$\Gamma^{i}_o$ ($\Gamma^{i}_e$) is proportional to a product of an odd (even)
number of gamma matrices. 
Gauge invariance requires that
\bqa
\label{eq:gaugeinv}
p_\alpha \int [d^4q]\cdots [d^4q_{\ell}] \bar J^\alpha(q,\cdots,q_{\ell}) = 0\,,  
\eqa
where $\bar J^\alpha$ is the integrand in Eq.~(\ref{eq:integg}) 
regulated {\em \`a la} FDR by replacing $q^2_i \to q^2_i-\mu^2$ in both numerators and denominators.
Eq.~(\ref{eq:gaugeinv}) can be directly proven at the integrand level. With this aim, we first concentrate on the replacements responsible for the conservation of the specific current in Fig.~\ref{genamp}:  
\bqa
 J^{\alpha} \to  J^{\prime \alpha} &=&  
 \sum_i \frac{1}{\bar D \bar D_p}
\Big(
     Tr \left[\rlap /\bar q \gamma^\alpha \rlap /\bar q  \Gamma^{i}_o \right]
    +Tr \left[\rlap /q \gamma^\alpha \rlap /p  \Gamma^{i}_o \right]
+m^2 Tr \left[\gamma^\alpha  \Gamma^{i}_o \right]  \nl
&& +m Tr \left[\gamma^\alpha (\rlap /q  + \rlap /p)  \Gamma^{i}_e \right]
+m Tr \left[\rlap /q  \gamma^\alpha   \Gamma^{i}_e \right]
\Big)
\,,
\eqa
where the loop denominators in  $\Gamma^{i}_{o,e}$ are also barred.
In the previous equation
\bqa
\label{repl}
\rlap /\bar q = \rlap /q \pm \mu\,
\eqa
has the effect of changing $q^2$ to $\bar q^2$ in the first trace. Thus, when contracting with $p$, it is possible to reconstruct and cancel denominators
\bqa
p_\alpha J^{\prime \alpha} &=& \sum_i  
\frac{1}{\bar D}
\Big(
  Tr   \left[ \rlap /q  \Gamma^{i}_o \right]
+ m Tr \left[\Gamma^{i}_e \right]
\Big)
\nl &&\;\;\;-
\frac{1}{\bar D_p}
\Big(
  Tr   \left[(\rlap /q + \rlap/p)  \Gamma^{i}_o \right]
+ m Tr \left[\Gamma^{i}_e \right]
\Big)\,,
\eqa
in agreement with the Feynman identity in Eq.~(\ref{feyd}). 
After that
\bqa
\label{eq:gaugeinvpr}
p_\alpha \int [d^4q]\cdots [d^4q_{\ell}] J^{\prime \alpha}(q,\cdots,q_{\ell}) = 0\,  
\eqa
directly follows from the shift invariance properties of the loop integrals, 
as in DR. We explicitly tested Eq.~(\ref{eq:gaugeinvpr}) up to two loops in $H \to \gamma \gamma$. 

With more photons, replacements as in Eq.~(\ref{repl}) have to be performed for all integration momenta appearing in the trace\,\footnote{Sums over internal indices have to be previously worked out.}. The one-loop prescription is that defined in~\cite{Donati:2013iya}:  given a fermionic string, one chooses arbitrarily the sign of $\mu$ within the first $\rlap / {\bar q}$; the sign of the subsequent one is opposite, if an even number of $\gamma$-matrices occur between the two $\rlap / {\bar q}$s, and it is the same otherwise\,\footnote{If chirality matrices are involved, a gauge invariant treatment requires their anticommutation at the beginning (or the end) of open strings before replacing 
$\rlap /q \to \rlap/\bar q$. In the case of closed loops, $\gamma_5$ should be put next to the vertex corresponding to a potential non-conserved current. This reproduces the correct coefficient of the triangular 
anomaly~\cite{Pittau:2012zd}.}.
This rule is sufficient in the presence of one fermion line only, as in the calculation at hand. With two or more lines, and no summation indices among them, each fermion string can be separately treated as described. If sums occur, after applying the above algorithm, extra $\mu^2$ terms need to be extracted according to the following procedure
\bqa
\label{eq:proced}
Tr \left[ ...\, \rlap /q \Gamma^{(n)} \gamma_\alpha\right] 
Tr \left[ ...\, \rlap /q \Gamma^{(m)} \gamma^\alpha\right] &\to& 
Tr \left[ ...\, \rlap /q \Gamma^{(n)} \gamma_\alpha\right] 
Tr \left[ ...\, \rlap /q \Gamma^{(m)} \gamma^\alpha\right] \nl
&-&(-1)^{(n+m)} \mu^2 
Tr \left[ ...\, \Gamma^{(n)}\right] 
Tr \left[ ...\, \Gamma^{(m)}\right],\nl
\eqa 
where $\Gamma^{(k)}$ represents a string of $k$ gamma matrices.
Eq.~(\ref{eq:proced}) is proven by noting that $n$ (m) anticommutation are needed to bring $\rlap /q$ near to $\gamma_\alpha$ ($\gamma^\alpha$) and can be easily checked by taking the traces and substituting $q^2 \to q^2 -\mu^2$.
As an example of such rules, the integrand of the one-loop 
$H \to \gamma(p_1) \gamma(-p_2) $ amplitude is proportional to
\bqa 
\label{gloex0}
J^{\alpha\beta}(q) &=& \frac{1}{D D_{p_1} D_{p_2}} 
Tr  \left[(\rlap /q +m ) \gamma^\alpha (\rlap /q + \rlap/p_1 +m ) (\rlap /q + \rlap/p_2 +m ) \gamma^\beta \right]\,,
\eqa
and its FDR regulated version reads
\bqa
\label{gloex}
\bar J^{\alpha\beta}(q) &=& 
\frac{1}{\bar D \bar D_{p_1} \bar D_{p_2}}
\Big( 
Tr  \left[(\rlap /q +m ) \gamma^\alpha (\rlap /q + \rlap/p_1 +m ) (\rlap /q + \rlap/p_2 +m ) \gamma^\beta \right]
\nl &&
+m \mu^2 Tr  \left[\gamma^\alpha \gamma^\beta \right]\Big)\,, 
\eqa 
which satisfies the Ward identities
\bqa
p_{1 \alpha} \int [d^4q] \bar J^{\alpha\beta}(q) = p_{2 \beta}  \int [d^4q] \bar J^{\alpha\beta}(q)  = 0\,.  
\eqa
We emphasize that there is nothing mysterious in Eq.~(\ref{gloex}): the same result is obtained by computing the trace in Eq.~(\ref{gloex0}) and replacing $q^2 \to \bar q^2$. The advantage of Eq.~(\ref{gloex}) is that it permits a trivial proof of the Ward identities at the integrand level.

The corresponding procedure at two loops is better explained with an example. Consider the trace
\bqa
\label{eq2lstart}
T^{\alpha \beta} =
Tr  \left[ 
\rlap /q_1 
\gamma^\alpha
\rlap /q_1 
\rlap /q_2 
\gamma^\beta
\rlap /q_2 
\right]\,,
\eqa
which contributes to the second diagram of Fig.~\ref{F_diagrams}.
Its FDR counterpart reads
\bqa
\label{eq2lo}
\bar T^{\alpha \beta} &=& T^{\alpha \beta} 
+ 
\mu^2|_1\, 
Tr  \left[ 
\gamma^\alpha
\rlap /q_2 
\gamma^\beta
\rlap /q_2 
\right]
+ 
\mu^2|_2\, 
Tr  \left[ 
\gamma^\alpha
\rlap /q_1 
\gamma^\beta
\rlap /q_1 
\right] \nl
&&+\mu^2|_1 \mu^2|_2
Tr  \left[ 
\gamma^\alpha
\gamma^\beta
\right]
-16  \tilde \mu^2_{12} q^\alpha_1 q^\beta_2\,, 
\eqa
with
\bqa
\tilde \mu^2_{12} = \frac{1}{2}
\Big(\mu^2|_{12}  -\mu^2|_{1}-\mu^2|_{2} \Big)\,.
\eqa
Eq.~(\ref{eq2lo}) is obtained from Eq.~(\ref{eq2lstart}) by using --\,one after the other\,-- the one-loop replacements $\rlap /q_1 \to \rlap /\bar q_1$ and $\rlap /q_2 \to \rlap / \bar q_2$, which generate the terms proportional to $\mu^2|_{1}$ and $\mu^2|_{2}$.
The $\tilde \mu^2_{12}$ contribution originates, instead, from the substitution
\bqa
(q_1 \cdot q_2) = \frac{1}{2}
\Big(q^2_{12} -q^2_{1} -q^2_{2} \Big) 
\to
\frac{1}{2}
\Big(\bar q^2_{12}-\bar q^2_{1}-\bar q^2_{2} \Big)\,,
\eqa 
and is obtained by simultaneously barring $\rlap /q_1$ and $\rlap /q_2$ in Eq.~(\ref{eq2lstart}) (with the same rule used at one loop to determine the sign of $\mu|_{i}$ inside each $\rlap /q_i$, without distinguishing between $\rlap /q_1$ and $\rlap /q_2$) and subtracting the $\mu^2|_i$ terms already calculated. What is left is, by construction, proportional to powers of $\mu|_{1}\mu|_{2} \equiv \tilde \mu^2_{12}$ and gives the last term in  
Eq.~(\ref{eq2lo}). 
Once again, $\bar T^{\alpha \beta}$ is equivalent to the replacements
\bqa
q^2_1 \to \bar q^2_1\,,~~
q^2_2 \to \bar q^2_2\,,~~
(q_1 \cdot q_2) \to
\frac{1}{2}
\Big(\bar q^2_{12}-\bar q^2_{1}-\bar q^2_{2} \Big)\,,
\eqa
in the original trace $T^{\alpha \beta}$. Thus, the 
$\mu^2|_{i}$ ensure the right cancellations leading to the fulfillment of the
Ward identities. We explicitly checked that the two-loop 
$H \to \gamma(p_1) \gamma(-p_2)$ integrand $\bar J^{\alpha\beta} (q_1,q_2)$, constructed as described, satisfies
\bqa
p_{1 \alpha} \int [d^4q_1] \int [d^4q_2] \bar J^{\alpha\beta}(q_1,q_2) = 
p_{2  \beta} \int [d^4q_1] \int [d^4q_2] \bar J^{\alpha\beta}(q_1,q_2) = 0\,.  
\eqa

In practical cases it is often convenient to simplify reducible numerators before computing the loop integrals. In that way, only irreducible tensors appear and extra integrals are just produced by tensor decomposition, as in Eq.~(\ref{eq:tensdec}). This is the strategy of the calculation presented in section~\ref{calculation}. 
%
%


\subsection{FDR versus DR}
\label{fdrvsdr}
The proof that DR preserves gauge invariance and unitarity relies on the possibility of introducing order-by-order local counterterms in the Lagrangian ${\cal L}$. On the contrary, FDR makes no reference to ${\cal L}$. In this subsection we use the simple two-loop QED example of~\cite{'tHooft:1973pz} to comment on the conceptual differences between the two approaches. 
%
%

Consider a DR calculation of the one-loop photon self-energy 
\vspace{1mm}
\begin{center}
\includegraphics[width=0.74\textwidth]{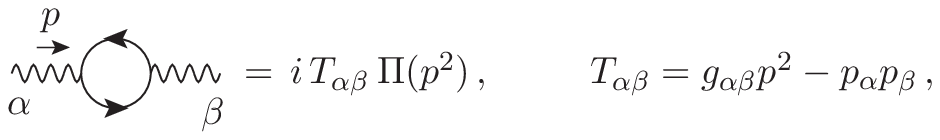}
\end{center}
with
\bqa
\Pi(p^2)     &=&  
\frac{1}{\epsilon}\, \Pi_{-1} + \Pi_{0} + \epsilon\, \Pi_{1} \,, \nl
\Pi_{0}  &=& \frac{e^2}{2 \pi^2} \int_0^1 dx \,x(1-x)\,
\ln \frac{m^2-p^2 x(1-x)}{\mur^2}\,.
\eqa
Then, at two loops and up to terms ${\cal O}(\epsilon)$,
\vspace{1mm}
\begin{center}
\includegraphics[width=\textwidth]{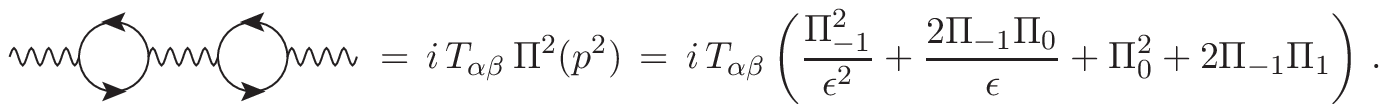}
\end{center}

Simply removing the poles from the last expression gives 
$\Pi_0^2 + 2\Pi_{-1}\Pi_1$,
which is not the right result because it violates unitarity. 
As is well known, the correct procedure to undertake in DR is to renormalize order by order, i.e. to add one-loop counterterms in  ${\cal L}$ such that
\vspace{2mm}
\begin{center}
\includegraphics[width=0.65\textwidth]{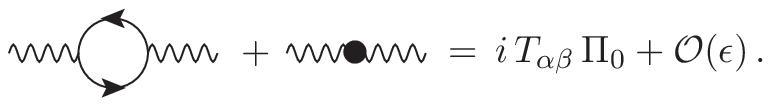}
\end{center}

Thus
\vspace{2mm}
\begin{center}
\includegraphics[width=\textwidth]{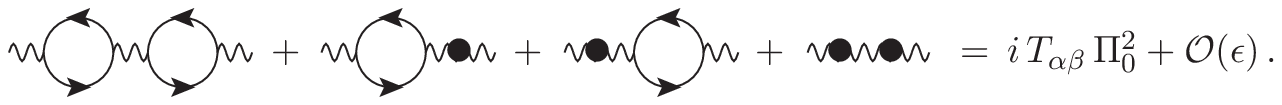}
\end{center}
\vspace{2mm}

In FDR, the divergences are subtracted at the level of the definition of the loop integration, so that the product of two one-loop diagrams is simply the product of the two finite parts, with no need of introducing extra interactions in  ${\cal L}$. Thus, one directly obtains
\vspace{2mm}
\begin{center}
\includegraphics[width=0.48\textwidth]{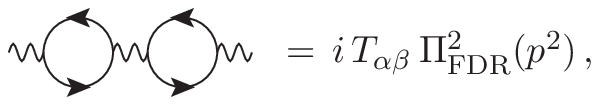}
\end{center}
with $\Pi_{\rm FDR}(p^2)= \Pi_0$.
This difference can be also understood from the {DR $\leftrightarrow$ FDR} naive correspondence
\bqa
\begin{tabular}{lll}
$\displaystyle  \epsilon$            & $\leftrightarrow$ & $\mu$ \\
$\displaystyle \frac{1}{\epsilon}$   & $\leftrightarrow$ & $\ln \mu\,$,
\end{tabular}
\eqa
which gives 
$\lim_{\epsilon \rightarrow 0} \epsilon/\epsilon = 1$,
while
$\lim_{\mu\rightarrow 0} \mu \ln\mu = 0$.

From all of that it is manifest that spurious $\epsilon/\epsilon$ terms such as $\Pi_{-1}\Pi_1$ --\,which need to be kept under control in DR by the order-by-order renormalization\,-- never appear in FDR.
The result of an FDR calculation typically depends on the parameters contained in ${\cal L}$, and a (finite) global renormalization is needed only to link them to experimental measurements at the desired perturbative accuracy. In particular --\,and in contrast with DR\,-- no renormalization is necessary when no parameter appears in the final result. This is the case of the calculation presented in section~\ref{calculation}.

\subsection{FDR versus FDH}
In this subsection we discuss the differences between FDR and the Four Dimensional Helicity scheme (FDH) of~\cite{Bern:2002zk}, which is equivalent to Dimensional Reduction~\cite{Siegel:1979wq} at one-loop. 
FDH is a variant of DR, in which gauge cancellations are kept by integrating 
all momentum integrals over $n$-component momenta and considering any $g^{\alpha \beta}$ resulting from the integration as $n$-dimensional.
Observed external states are treated in four dimensions (preserving supersymmetry) and unobserved internal ones are defined in such a way that the contraction $q^{\alpha} q^{\beta} g_{\alpha \beta}= q^2$ gives rise to an $n$-dimensional object when $q$ is an integration momentum.
If $q^2$ is split into four-dimensional ($q^2_4$) and $\epsilon$-dimensional
($\tilde q^2$) components 
\bqa
q^2= q^2_4 + \tilde q^2\,,
\eqa 
and $\tilde q^2$ is identified with $-\mu^2$, there is a formal equivalence 
--\,at the integrand level\,-- between the procedures used  by FDR and FDH to determine the $\mu^2$ pieces~\cite{Bern:1995db,Cullen:2011ac}. However differences start when integrating.
FDR integration is defined in a way that non-local sub-divergences are subtracted right away, as in the second line of Eq.~(\ref{eq:ex2}), while in FDH sub-divergences are compensated by counterterms added at a previous renormalization 
stage, as in any conventional subtraction scheme. It is exactly this peculiarity that makes possible to avoid an order-by-order renormalization in FDR.

As a consequence of this dissimilarity, integrals containing $\mu^2$ give different results, at two loops and beyond, when computed in FDR and FDH.
For example
\bqa
\int [d^4q_1][d^4q_2] \frac{\mu^2|_1}{(\bar q^2_1-m^2)^2
(\bar q^2_1-m^2)^2(\bar q^2_{12}-m^2)}= 
\pi^4
\left(
\frac{2}{3} f + \frac{1}{2} \ln \frac{m^2}{\mur^2}
\right), 
\eqa
with $f$ defined in Eq.~(\ref{eq:f}), while
\bqa
\int d^nq_1d^nq_2 \frac{-\tilde q^2_1}{(q^2_1-m^2)^2
(q^2_1-m^2)^2(q^2_{12}-m^2)} &=& 
\pi^4
\Big(\frac{1}{2 \epsilon}-\frac{3}{8} + \frac{1}{2} \ln \frac{m^2}{\mur^2}\nl
&&+ \frac{\gamma_E+ \ln \pi}{2} \Big) + {\cal O}(\epsilon)\,. 
\eqa
Only at one loop, because no sub-divergences are present, FDR and FDH coincide, as observed in~\cite{Pittau:2013qla}. For instance,
\bqa
\int [d^4q]       \frac{\mu^2}{(q^2-M^2)^3}=
\int  d^nq  \frac{-\tilde q^2}{(q^2-M^2)^3}= \frac{i \pi^2}{2}\,.
\eqa

\subsection{Infrared divergences}
\label{ircl}
Although the process $H \to \gamma \gamma$ is free of IR infinities, we devote this subsection to illustrate how soft and collinear singularities can be  treated compatibly with FDR. We first discuss divergences in the virtual contribution, and then show how they are matched by a particular treatment of the real radiation. 
\begin{figure}[h]
\begin{center}
\includegraphics[width=0.8\textwidth]{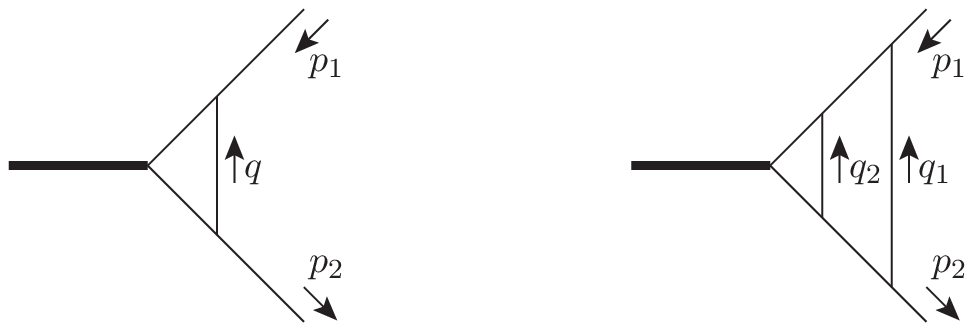}
\end{center}
\caption{ \small Examples of massless one-loop and two-loop scalar integrals.
Thin lines represent massless scalar propagators and $p^2_1=p^2_2= 0$.
} \label{massless}
\end{figure}

As for the loop integration, the definition in Eq.~(\ref{eq:fdrdef}) can be maintained also in the presence of IR singularities.
For instance, the FDR versions of the massless one- and two-loop scalar integrals in Fig.~\ref{massless} read
\bqa
\int[d^4q]\,J^{(1)}(q,\mu^2) &=&
\lim_{\mu \to 0}
\int d^4q\, J^{(1)}(q_1,\mu^2)\,~~~{\rm and} \nl
\int[d^4q_1][d^4q_2]\, J^{(2)}(q_1,q_2,\mu^2) &=&
\lim_{\mu \to 0}
\int d^4q_1 d^4q_2\, J^{(2)}(q_1,q_2,\mu^2)\,,
\eqa
respectively, with\,\footnote{$J^{(1)}$ and $J^{(2)}$ are not UV subtracted since they produce UV convergent integrals.} 
\bqa
J^{(1)}(q,\mu^2)       &=& \frac{1}{\dbar_0(q) \dbar_{p_1}(q) \dbar_{p_2}(q)}\,,\nl 
J^{(2)}(q_1,q_2,\mu^2) &=& J^{(1)}(q_1,\mu^2)\, \frac{1}{ 
\dbar_0(q_2) \dbar_{p_1}(q_{12}) \dbar_{p_2}(q_{12})}\,, \nl
\dbar_{p_i}(q_j) &=& \qbar^2_j+2 (q_j \cdot p_i)\,.
\eqa
Note that the on-shell conditions $p^2_1= p^2_2= 0$ are used at the integrand level. Thus, infrared virtual divergences get regulated by the $\mu^2$-deformed  propagators\,\footnote{A different $\mu^2$ can be used to regulate UV divergences ($\mu^2_{\rm UV}$) and IR ones ($\mu^2_{\rm IR}$). However, a common  $\mu^2$ simplifies the calculation, as will be shown later. Since IR infinities are more easily understood in terms of  $\mu^2_{\rm IR} > 0$, it is convenient to choose $\mu^2_{\rm UV}= \mu^2_{\rm IR}= \mu^2 > 0$.}, which generates powers of logarithms of $\mu^2$, upon integration.
A particularly interesting situation is when the integral is also UV divergent. In this case it is easy to see that UV divergent scale-less $\ell$-loop FDR integrals vanish, as in DR.
In fact, the only allowed external variable is a momentum $p$ such that $p^2=0$,  whose fate is to appear in the numerator of $J_{{\rm F},\ell}(q_1,\ldots, q_\ell)$ in Eq.~(\ref{eq:fdrexpansion}) to improve the UV convergence of the original integrand.
Therefore, $J_{{\rm F},\ell}(q_1,\ldots, q_\ell)$ is entirely made of integrands proportional to positive powers of $(q_i \cdot p)$, that vanish, by Lorentz invariance, after integration.
The simplest case is the fully massless one-loop 2-point scalar function
\bqa
B^{\rm FDR}(p^2=0,0,0) = \int [d^4q]\frac{1}{\qbar^2((q+p)^2-\mu^2)}
\,.
\eqa
The FDR expansion of its integrand reads
\bqa
\frac{1}{\qbar^2 \dbar_p} =
\left[\frac{1}{\qbar^4} \right]
-2 \frac{(q \cdot p)}{\qbar^4 \dbar_p}\,,
\eqa
so that
\bqa
\label{eq:uvcl}
B^{\rm FDR}(p^2=0,0,0) =  -2  \lim_{\mu \to 0} \int d^4q
\frac{(q \cdot p)}{\qbar^4 \dbar_p} = 0\,.  
\eqa
The same result is obtained by a direct computation
\bqa
B^{\rm FDR}(p^2,0,0) = 
 -i \pi^2 \lim_{\mu \to 0}\int_0^1 dx\, \left(
\ln(\mu^2 -p^2 x (1-x))-\ln(\mu^2) \right)\,,
\eqa
from which it is manifest that, in the limit $p^2 \to 0$, a cancellation occurs
between two logarithms of UV and IR origin, respectively\,
\footnote{
\label{foot1}
It is instructive to study the same case in DR, where
$
B^{\rm DR}(p^2,0,0) = \mur^{-\epsilon} \int d^nq \frac{1}{q^2{(q+p)^2}}
$.
Now $B^{\rm DR}(0,0,0)$ vanishes because IR and UV poles in $\epsilon$ compensate. In fact, by introducing an arbitrary separation scale $M$, the two divergences can be disentangled
\bqa
{ \frac{1}{(q+p)^2}}= \frac{1}{q^2-M^2} 
-\bigg( 
 \frac{1}{q^2-M^2}
-\frac{1}{(q+p)^2} \bigg) 
= {\frac{1}{q^2-M^2}}-{\frac{M^2+2 (q \cdot p)}{(q^2-M^2)(q+p)^2}}\,.
\eqa
Then the integrals ($\label{Delta}
\Delta= -\frac{2}{\epsilon}-\gamma_{E} - \ln \pi $)
\bqa
I_{\rm UV} &=&  \mur^{-\epsilon} \int d^nq \frac{1}{q^2 (q^2-M^2)} =  i \pi^2
\left(
\Delta -\ln\frac{M^2}{\mur^2} +1
\right)\,,
\nl
I_{\rm IR} &=& \mur^{-\epsilon} \int d^nq \frac{M^2+2 (q \cdot p)}{q^2(q^2-M^2)(q+p)^2} = I_{\rm UV}\,
\eqa
cancel each other.
However, this argument has a potential problem, because it requires the cancellation of two analytic continuations, $I_{\rm UV}$ and $I_{\rm IR}$, originally defined in domains that {\em do not overlap}~\cite{Leibbrandt} ($\epsilon < 0$ and $\epsilon > 0$): since no value of $\epsilon$ exists where they are defined
simultaneously, it is not obvious whether their difference represents the original function $B^{\rm DR}(0,0,0)$. A possible mathematically consistent solution can be formulated in terms of modified Gaussian integrals in the $n$-dimensional Euclidean space~\cite{Leibbrandt}. In contrast, the FDR derivation in Eq.~(\ref{eq:uvcl}) is straightforward.}. 

In summary, IR divergent loop integrals are defined by taking the limit $\mu \to 0$ outside integration, after subtracting --\,when necessary\,-- UV divergent integrands. In order to preserve the cancellation of the IR logarithms in physical quantities, this definition should be accompanied by a consistent treatment of the infinities appearing in the real emission, which we discuss in the following. 

\begin{figure}[t]
\begin{center}
\includegraphics[width=.6\textwidth]{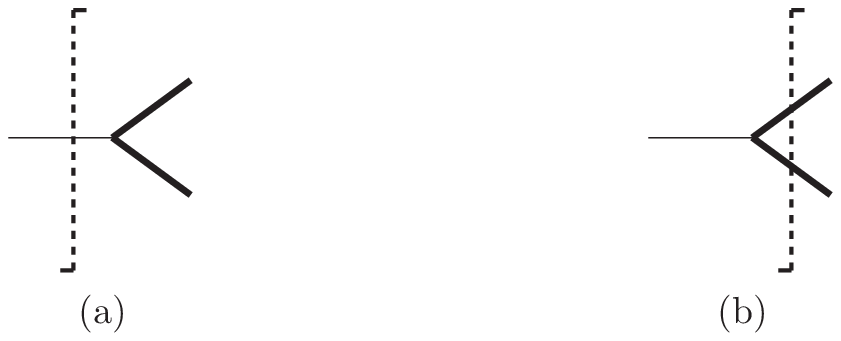}
\caption{\label{fig:fig2} Splitting regulated by massive (thick) unobserved particles. The one-particle cut in (a) contributes to the virtual part, the two-particle cut of (b) to the real radiation.}
\end{center}
\end{figure}

Consider how the divergent $1 \to 2$ splitting is regulated in the loop integrals. The situation is depicted in Fig.~\ref{fig:fig2}(a), where thick lines represent unobserved loop particles --\,whose propagator is made massive by the addition of $\mu^2$\,-- and the cut line is an external observed massless particle.
This is matched by the real radiation pattern of Fig.~\ref{fig:fig2}(b), where
thick lines are unobserved external particles merging into an internal observed massless one.
In both situations unobserved particles get a mass $\mu$ and unitarity relates the two cases as follows
\bqa
\frac{1}{q^2-\mu^2} \leftrightarrow \delta(q^2-\mu^2)\, \theta(q(0))\,.
\eqa

Therefore, would-be-massless external particles with momenta $p_i$ should be given a mass $\mu$. This is achieved by trading the original massless m-body phase space $d \Phi_m$ for a massive one, denoted by $d \bar \Phi_m$, in which 
\bqa
\label{repm}
p_i^2 \to \bar p^2_i = \mu^2\,. 
\eqa
In this way, singular configurations produce logarithms which cancel the IR dependence on $\mu^2$ of the virtual contribution. However, this strategy should be carried out without breaking gauge invariance. To illustrate the way to proceed we consider $m$-jet production at NNLO in $e^+ e^-$ annihilation. 
The building blocks of the calculation depend on the set of invariants 
\bqa
\{s_{i_1 \div i_{m}}\} \equiv \{s_{i_1 i_2}, s_{i_1 i_2 i_3}, \ldots, s_{i_1 \cdots i_{m}} 
\},~s_{i \cdots j} = (p_i + \cdots + p_j)^2,~p^2_i = 0.
\eqa 
They are: 
\begin{itemize}
\item the Born contribution $d \sigma^{B}_{\rm LO}\{s_{i_1 \div i_{m-1}}\}$, 
\item the virtual and real NLO corrections, $d \sigma^{V}_{\rm NLO}\{s_{i_1 \div i_{m-1}}\}$ and $d \sigma^{R}_{\rm NLO}\{s_{i_1 \div i_{m}}\}$,
\item the NNLO two-loop part $d \sigma^{V,2}_{\rm NNLO}\{s_{i_1 \div i_{m-1}}\}$,
\item the one-loop corrections to the NLO real radiation, $d \sigma^{V,1}_{\rm NNLO}\{s_{i_1 \div i_{m}}\}$, 
\item the double radiation $d \sigma^{R}_{\rm NNLO}\{s_{i_1 \div i_{m+1}}\}$. 
\end{itemize}
After $\alpha_S$ renormalization, they give a $m$-jet cross section accurate up to NNLO
\bqa
d \sigma &=& d \sigma_{\rm LO} + d \sigma_{\rm NLO} + d \sigma_{\rm NNLO}\,, 
\eqa
where
\bqa
\label{eqlnll}
d \sigma_{\rm LO}  &=& \int_{d \Phi_m} d \sigma^{B}_{\rm LO}\{s_{i_1 \div i_{m-1}}\}\,, \nl
d \sigma_{\rm NLO} &=& 
  \int_{d \Phi_m}    d \sigma^{V}_{\rm NLO}\{s_{i_1 \div i_{m-1}}\} 
+ \int_{d \Phi_{m+1}} d \sigma^{R}_{\rm NLO}\{s_{i_1 \div i_{m}}\}\,, \nl
d \sigma_{\rm NNLO} &=&
 \int_{d \Phi_m}    d \sigma^{V,2}_{\rm NNLO}\{s_{i_1 \div i_{m-1}}\}
+\int_{d \Phi_{m+1}} d \sigma^{V,1}_{\rm NNLO}\{s_{i_1 \div i_{m}}\}
\nl
&&+\int_{d \Phi_{m+2}} d \sigma^{R}_{\rm NNLO}\{s_{i_1 \div i_{m+1}}\}\,.
\eqa
The integrands behave as 
\bqa
d \sigma\{\cdots \} \sim \frac{1}{s_{ij}}  
\,,   ~~{\rm if}~~s_{ij} \to 0~~~~~{\rm and}~~~~~ 
d \sigma\{\cdots \} \sim \frac{1}{s^2_{ijk}}\,,~~{\rm if}~~s_{ijk} \to 0 \,,
\eqa
therefore the integrations over single- and double-unresolved massless phase-spaces ($\int_{d \Phi_{m+1}}$ and $\int_{d \Phi_{m+2}}$, respectively) generate logarithmic IR divergences which have to be regulated.
In DR, the last two lines of Eq.~(\ref{eqlnll}) are interpreted as a limit
to $\epsilon \to 0$ of integrals computed in $n= 4 + \epsilon$ dimensions. 
We instead define a mapping from massless to massive invariants as follows
\bqa
\label{mapping}
s_{i_1 \cdots i_m} &\to& \hat s_{i_1 \cdots i_m} \equiv \sum_{k < l}^m   \hat s_{i_k i_l}\,, \nl
\hat s_{ij}  &=& \bar s_{ij} = (\bar p_i+\bar p_j)^2 \,, \nl
\bar p^2_i  &=& \mu^2\,,
\eqa
and rewrite
\bqa
\label{eqmast}
d \sigma_{\rm NLO} &=&  
 \int_{d \Phi_m}    d \sigma^{V}_{\rm NLO}\{s_{i_1 \div i_{m-1}}\} 
+ \lim_{\mu \to 0} 
 \int_{d \bar \Phi_{m+1}} d \sigma^{R}_{\rm NLO}\{\hat s_{i_1 \div i_{m}}\}\,, \nl
d \sigma_{\rm NNLO} &=& 
 \int_{d \Phi_m}    d \sigma^{V,2}_{\rm NNLO}\{s_{i_1 \div i_{m-1}}\}
+ \lim_{\mu \to 0} \int_{d \bar \Phi_{m+1}} d \sigma^{V,1}_{\rm NNLO}\{\hat s_{i_1 \div i_{m}}\} \nl
&&+ \lim_{\mu \to 0} \int_{d \bar \Phi_{m+2}} d \sigma^{R}_{\rm NNLO}\{\hat s_{i_1 \div i_{m+1}}\} \,W_{\rm NNLO} \{\hat s_{i_1 i_2 i_3}\}_{m+1}\,,
\eqa
where $\mu$ is the same regulator used in the IR divergent loop integrals, and
\bqa
\label{weight}
W_{\rm NNLO} \{\hat s_{i_1 i_2 i_3}\}_{m+1}= \prod_{i<j<k}^{m+1} 
\left(\frac{\hat s_{ijk}}{\bar s_{ijk}}
\right)^2\,,~~~~~\bar s_{ijk}= (\bar p_i+ \bar p_j+ \bar p_k)^2\,.
\eqa 

The proof that Eq.~(\ref{eqmast}) converges to the right results is simple.
First note that the mapping in Eq.~(\ref{mapping}) preserves all formal properties of massless kinematics. For instance
\bqa
\hat s_{123}= \hat s_{12} + \hat s_{13} + \hat s_{23}\,.
\eqa
Thus, $d \sigma^{R}_{\rm NLO}$, $d \sigma^{V,1}_{\rm NNLO}$ and $d \sigma^{R}_{\rm NNLO}$ are gauge invariant  by construction.
As for the NLO real emission, $\frac{1}{s^2_{ijk}}$ poles are always screened by the requirement of observing $m$ particles. Therefore, the only possible singular behavior is
\bqa
d \sigma^{R}_{\rm NLO}\{\hat s_{i_1 \div i_{m}}\}  &\sim& 
 \frac{1}{\hat s_{ij}}=
 \frac{1}{\bar s_{ij}} =  \frac{1}{(\bar p_i+\bar p_j)^2}\,,
\eqa
which, being the internal propagator massless, matches the virtual IR poles,
as in Fig.~\ref{fig:fig2}(b).
In the NNLO case, $d \sigma^{R}_{\rm NNLO}\{\hat s_{i_1 \div i_{m+1}}\}$ contains
additional $\frac{1}{\hat s^2_{ijk}}$ poles, which no longer have the form of massless propagators. In fact, a spurious mass is generated by the gauge invariant mapping of Eq.~(\ref{mapping}): 
\bqa
\hat s_{ijk}= \bar s_{ijk}+ 3 \mu^2= (\bar p_i+\bar p_j +\bar p_k)^2+ 3 \mu^2\,. 
\eqa  
To cure this, $d \sigma^{R}_{\rm NNLO}$ is multiplied by the weight factor in Eq.~(\ref{weight}), which changes --\,in a gauge invariant way\,--  any pole 
$
\frac{1}{\hat s^2_{ijk}} 
$
to the correct value
$
\frac{1}{\bar s^2_{ijk}}\,.
$
The additional integrals, generated when each term in $W_{\rm NNLO}$ does not 
meet its corresponding pole, vanish in the limit $\mu \to 0$.
This last property follows from the fact that the integral is at most logarithmically divergent.
The reason why a pole cannot be changed {\em by hand} only in the terms where it appears, is that the logarithmic behavior is reached only {\em after} gauge cancellations, which should not be altered. This can be easily understood with a toy model: 
\bqa
\left. d \sigma^{R}_{\rm NNLO}\right|_{toy}= \frac{1}{\hat s^2_{12}} 
-\frac{( \hat s_{12}+\hat s_{13}+ \hat s_{23})^2}{\hat s^2_{12} 
\hat s^2_{123}}\,.
\eqa
The correct procedure gives
\bqa
\lim_{\mu \to 0} \int_{d \bar \Phi_{m+2}} \left. d \sigma^{R}_{\rm NNLO}\right|_{toy}\,W_{\rm NNLO} \{\hat s_{i_1 i_2 i_3}\}_{m+1} = 0\,,
\eqa
while
\bqa
\lim_{\mu \to 0} \int_{d \bar \Phi_{m+2}}
\left[\frac{1}{\hat s^2_{12}} 
-\frac{( \hat s_{12}+\hat s_{13}+ \hat s_{23})^2}{\hat s^2_{12} 
\bar s^2_{123}}\right] \ne 0\,.
\eqa

To summarize, IR infinities can be safely treated in four dimensions. An explicit one-loop example, involving both IR and UV infinities, can be found in~\cite{Pittau:2013qla}. Furthermore, the outlined strategy opens the possibility of a numerical treatment of NNLO calculations similar to the phase-space slicing method at NLO~\cite{Giele:1993dj}. The advantage is that all singularities are automatically expressed in terms of powers of a logarithmic regulator --\,$\ln \mu$\,-- with no need of subtracting $1/\epsilon$ poles. An investigation of the numerical performance of such a strategy is outside the scope of this work, although currently under study. We think that it is a promising one because, owing to the four-dimensionality of the calculation, we envisage that the bulk of the cancellations can be easily arranged to happen at the integrand level.

\section{$\mathbf H \to \gamma \gamma$ at two loops} 
\label{calculation}
\begin{figure}
\begin{center}
\includegraphics[width=0.9\textwidth]{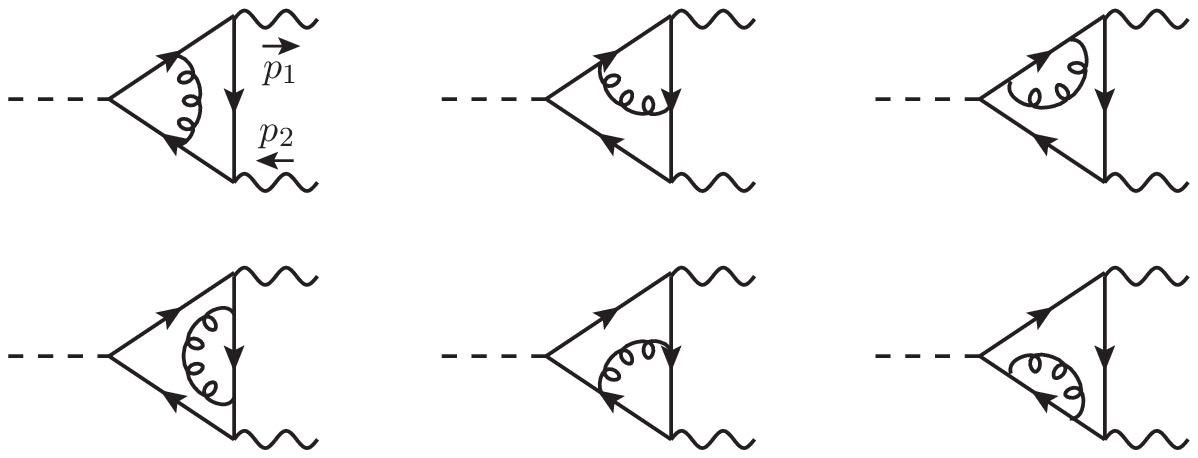}
\end{center}
\caption{ \small Feynman diagrams contributing to the QCD corrections of the top-loop-mediated Higgs decay into two photons. The same diagrams with the electric charge flowing counterclockwise also contribute.} \label{F_diagrams}
\end{figure}

The diagrams contributing to the QCD corrections of the top-loop-mediated Higgs decay into two photons are depicted in Fig.~\ref{F_diagrams}.

The amplitude reads
\bqa
\label{eq:amp}
	\M = \M^{\alpha\beta} \eps_{\alpha}(p_1) \eps_{\beta}^*(p_2)\,,
\eqa
where $p_1$ and $-p_2$ are the momenta of the outgoing photons.
One has
\bqa
	\M^{\alpha\beta} = \frac{1}{(2\pi)^2}\frac{\alpha}{\pi}
			        	\frac{T^{\alpha\beta}}{v} 
\frac{4}{3}\eta \F(\eta)\,,
\eqa
with $v$ being the \emph{vev} of the Higgs boson and
\bqa
 \eta= \frac{4m^2}{s}\,,~~ m = m_{\rm top}\,,~~ s=(p_1-p_2)^2= M^2_H\,,~~
T^{\alpha\beta} = p_2^{\alpha}p_1^{\beta} + \frac{s}{2} g^{\alpha\beta}\,.
\eqa
$\M$ is well defined in the limit $m \to \infty$ we are interested in. This means that, order by order, the form factor $\F(\eta)$ can be written as
\bqa
\label{eq:etaexp}
	\F(\eta) =   
		\F_0 +  \frac{\F_1}{\eta} +  \frac{\F_2}{\eta^2} +  \ldots
	\,
\eqa
with
\bqa
\F_0= 0\,.
\eqa
By inserting Eq.~(\ref{eq:etaexp}) into the expansion in $\alpha_S$ of $\F(\eta)$, one obtains, up to two loops and neglecting $\ord(\tfrac{1}{\eta^2})$ terms, 
\bqa
\F(\eta) &=&
	\F^{(1)}(\eta) 
		-i \frac{\alpha_S}{3 \pi^3} 
		  \F^{(2)}(\eta) \nl 
&\equiv&
        \left(\F^{(1)}_0 +  \frac{\F^{(1)}_1}{\eta}\right) 
		-i \frac{\alpha_S}{3 \pi^3} 
        \left(\F^{(2)}_0 +  \frac{\F^{(2)}_1}{\eta}\right)\,. 
\eqa
At one loop $\F^{(1)}_0 = 0$ and (see, for example,~\cite{Donati:2013iya}) 
\bqa
\F^{(1)}_1 = \frac{4 i \pi^2}{3}\,.
\eqa
In this section, we re-derive\,\footnote{We use the Feynman rules of appendix~\ref{app0}.} --\,within the FDR framework\,-- the known result~\cite{Djouadi:1990aj}
\bqa
\label{eq:expected}
\F^{(2)}_0 &=& 0\,, \nl
\F^{(2)}_1 &=& 4 \pi^4\,,
\eqa
which implies that the QCD corrections factorize the one-loop amplitude
\bqa
\M = \M^{(1)}\left(1-\frac{\alpha_S}{\pi}\right) 
+ {\cal O}\left(\alpha^2_S\right)
+ {\cal O}\left(\frac{1}{\eta}\right)\,.
\eqa

\subsection{The building blocks}
\label{sec:bblo}
Since we are working in the large top mass limit, denominators can be expanded 
as follows
\begin{align}
	\frac{1}{(q_i+p_j)^2-m^2} 
	& = 
	\frac{1}{q_i^2-m^2}\Bigg( 
		1-\frac{2(q_i\cdot p_j)+p_j^2}{(q_i+p_j)^2-m^2}
	\Bigg)\, \nonumber
	\\
	& = \frac{1}{q_i^2-m^2}\Bigg( 
		1-\frac{2(q_i\cdot p_j)}{q_i^2-m^2} \,+\, \ldots
	\Bigg)\,,
\end{align}
where the on-shell condition $p_j^2=0$ for the photons has been used. An expansion to the second order, as the one above, is sufficient to the level of accuracy we are interested in, i.e. $\ord(1/\eta)$. 
All external momenta can then be neglected and the top mass is the only relevant scale. As a consequence, we only have to deal with vacuum integrals.

After cancelling between numerator and denominator the $\qbar^2_1$, $\qbar^2_2$, $\qbar^2_{12}$ terms generated by the Feynman rules\,\footnote{Remember the discussion at the beginning of subsection \ref{subsec:gauge}.}, tensor integrals up to rank 4 contribute to the amplitude. Because there is no dependence on external momenta, odd rank integrals vanish and the tensor reduction gives:
\begin{align}
	q_i^{\alpha}q_j^{\beta}
	& \rightarrow \frac{(q_i\cdot q_j)}{4} g^{\alpha\beta} \,, \nonumber
	\\
	q^{\alpha}q^{\beta}q^{\rho}q^{\sigma}
	& \rightarrow \frac{q^4}{24}g^{\alpha\beta\rho\sigma}
	~~\text{at~one~loop\,,} \nonumber
	\\
	q_a^{\alpha}q_b^{\beta}q_r^{\rho}q_s^{\sigma}
	& \rightarrow \frac{1}{72} 
	\Big(A_{abrs}^{\alpha\beta\rho\sigma}+A_{arbs}^{\alpha\rho\beta\sigma}+A_{asbr}^{\alpha\sigma\beta\rho} \Big)~~\text{at~two~loops\,,}
\end{align}
where $g^{\alpha\beta\rho\sigma}= g^{\alpha\beta}g^{\rho\sigma}+g^{\alpha\rho}g^{\beta\sigma}+
g^{\alpha\sigma}g^{\beta\rho}$, and  
\bqa
A_{abrs}^{\alpha\beta\rho\sigma} = \Big( 5 (q_a\cdot q_b)(q_r\cdot q_s)-(q_a\cdot q_r)(q_b\cdot q_s)-(q_a\cdot q_s)(q_b\cdot q_r)\Big) g^{\alpha\beta} g^{\rho\sigma}\,.\nl
\eqa
Denominators can then be reconstructed by rewriting
\bqa
q^2_1 &=&  \qbar^2_{1}+\mu^2|_1\,,~~
q^2_2 = \qbar^2_{2}+\mu^2|_2\,,~~ \nl
2(q_1\cdot q_2) &=&  \qbar^2_{12}-\qbar^2_{1}-\qbar^2_{2}
		+\mu^2|_{12}-\mu^2|_{1}-\mu^2|_{2}\,.
\eqa
During this tensor decomposition, the $\mu^2|_1$, $\mu^2|_2$, $\mu^2|_{12}$ terms
are kept only when they generate a non-zero contribution. This means that they should be power-counted as the corresponding squared loop momenta, and contribute only if the integral is divergent.
The final result can then be completely expressed in terms of scalar two-loop integrals, products of two one-loop integrals and extra integrals containing $\mu^2|_j$ ($j= 1,2,12$). For convenience, we introduce the notation
\begin{align}
\label{eq:scalarint1}
	\big[ \,\alpha m \,\,\big] 
	& = \int \frac{\fdr}{(\qbar^2-m^2)^{\alpha}}\,,
        \\
\label{eq:scalarint12}
	\big[ \,\alpha m_1 \,|\, \beta m_2 \,\big] 
	& = 
	\int \frac{\fdru}{(\qbar_1^2-m^2_1)^{\alpha}}
	\times
	\int\frac{\fdrd}{(\qbar_2^2-m^2_2)^{\beta}}\,, 
	\\  
\label{eq:scalarint2}
	\big[ \,\alpha m_1\,|\, \beta m_2\,|\, 0 \,\big]
	& = 
	\int \frac{\fdru\fdrd}{(\qbar^2_1-m^2_1)^{\alpha}(\qbar^2_2-m^2_2)^{\beta}\qbar^2_{12}}\,,
\end{align}
and
\begin{align}
\label{eq:extraint}
	\big[ \,\alpha m \,\,\big](\mu^2) 
	& = \int \frac{\fdr\,\mu^2}{(\qbar^2-m^2)^{\alpha}}\,,
        \\
	\big[ \,\alpha m_1 \,|\, \beta m_2 \,\big] (\mu^2|_1)
	& = 
	\int \frac{\fdru\,\mu^2|_1}{(\qbar_1^2-m^2_1)^{\alpha}}
	\times
	\int\frac{\fdrd}{(\qbar_2^2-m^2_2)^{\beta}} \,, \nonumber
	\\  
	\big[ \,\alpha m_1\,|\, \beta m_2\,|\, 0 \,\big](\mu^2|_j)
	& = 
	\int \frac{\fdru\fdrd\, \mu^2|_j}{(\qbar^2_1-m^2_1)^{\alpha}(\qbar^2_2-m^2_2)^{\beta}\qbar^2_{12}}\,.
\end{align}
The one-loop and factorizable integrals of Eqs.~(\ref{eq:scalarint1}) and~(\ref{eq:scalarint12}) can be computed as derivatives of the quadratically divergent one-loop tadpole~\cite{Pittau:2012zd}
\bqa
	\int\fdr \frac{1}{(\qbar^2-m^2)^{\alpha}} &=&
	\frac{1}{\Gamma(\alpha)} 
 	\frac{\de^{\alpha-1}}{\de (m^2)^{\alpha-1}}\,\int\fdr \frac{1}{(\qbar^2-m^2)}\,, 
\nl
\int\fdr\frac{1}{(\qbar^2-m^2)} 
&=& -i\pi^2\,m^2\bigg(\log\frac{m^2}{\mur^2}-1 \bigg)\,,
\eqa 
while those  in Eq.~(\ref{eq:scalarint2}) are obtained by deriving with respect to the mass parameters the basic integral 
\bqa
\label{eq:eqfunint}
\big[ \,2 m_1\,|\, m_2\,|\, 0 \,\big]
\eqa
computed in appendix~\ref{appb}\,\footnote{This implies that for each of the diagrams in Fig.~\ref{F_diagrams} the routing of the momenta is chosen such as the gluon line gets the momentum $q_{12}$. This is allowed due to the shift invariance properties of the FDR integration.}
\bqa
	\big[\,\alpha\, m_1 \,|\, \beta\, m_2 \,| \, 0 \,\big]
	= \frac{1}{\Gamma(\alpha)\Gamma(\beta)}
	\frac{ \de^{\alpha-2}}
	{\de (m_1^2)^{\alpha-2}}
	\frac{ \de^{\beta-1}}
	{\de (m_2^2)^{\beta-1}}
\big[ \,2 m_1\,|\, m_2\,|\, 0 \,\big]\,.
\eqa
All extra integrals relevant for our calculation can be expressed in terms of three fundamental objects
\bqa
\mu^2 \int d^4q \frac{1}{\qbar^6} &=& -\frac{i\pi^2}{2}\,, \nl
\mu^2\int d^4q_1 d^4q_2
	\frac{q_1^2+2(q_1\cdot q_2)}{\qbar_1^6\qbar_2^4\qbar^2_{12}} 
&=& 
	-\pi^4 \left(\frac{2}{3} f +\frac{1}{2} \right)\,, \nl
\mu^2\int d^4q_1 d^4q_2 
\frac{1}{\qbar_1^4\qbar_2^4\qbar^2_{12}} 
&=& -\frac{2\,\pi^4}{3} f\,,
\eqa
with $f$ given in Eq.~(\ref{eq:f}).
We need 
\bqa
\big[ \,3 m\,|\, m\,|\, 0 \,\big](\mu^2|_1)\,,
\eqa
derived in footnote~\ref{foot}, and 
\bqa
	\int\fdr \frac{\mu^2}{(\qbar^2-m^2)^{\alpha}} &=&
	\frac{1}{\Gamma(\alpha)} 
 	\frac{\de^{\alpha-1}}{\de (m^2)^{\alpha-1}}\,\int\fdr \frac{\mu^2}{(\qbar^2-m^2)} \,,\nl
	\big[\,\alpha\, m_1 \,|\, \beta\, m_2 \,| \, 0 \,\big](\mu^2|_j)
	&=& \frac{1}{\Gamma(\alpha)\Gamma(\beta)}
	\frac{ \de^{\alpha-2}}
	{\de (m_1^2)^{\alpha-2}}
	\frac{ \de^{\beta-2}}
	{\de (m_2^2)^{\beta-2}}
\big[ \,2 m_1\,|\, 2 m_2\,|\, 0 \,\big](\mu^2|_j)\nl
\eqa
with
\bqa
\label{eq:scalext}
\int\fdr \frac{\mu^2}{(\qbar^2-m^2)} &=& 
- m^4 \lim_{\mu \to 0} \mu^2 \int d^4q \frac{1}{\qbar^6}\,,
\nl
\big[ \,2 m_1\,|\, 2 m_2\,|\, 0 \,\big](\mu^2|_1) 
&=& 
-\lim_{\mu \to 0} \mu^2 \left\{\int 
\frac{d^4q_1 d^4q_2}{\qbar_1^4\qbar_2^4\qbar^2_{12}}
-i \pi^2 \ln\frac{m^2_2}{\mu^2} \int 
\frac{d^4q}{\qbar^6} 
\right\},
\nl 
\big[ \,2 m_1\,|\, 2 m_2\,|\, 0 \,\big](\mu^2|_2) 
&=& 
-\lim_{\mu \to 0} \mu^2 \left\{\int 
\frac{d^4q_1 d^4q_2}{\qbar_1^4\qbar_2^4\qbar^2_{12}}
-i \pi^2 \ln\frac{m^2_1}{\mu^2} \int 
\frac{d^4q}{\qbar^6} 
\right\},
\nl
\big[ \,2 m_1\,|\, 2 m_2\,|\, 0 \,\big](\mu^2|_{12}) &=&  
-\lim_{\mu \to 0} \mu^2 \left\{\int 
\frac{d^4q_1 d^4q_2}{\qbar_1^4\qbar_2^4\qbar^2_{12}}
\right. \nl &&- \left.
i \pi^2 \left(\ln\frac{m^2_1}{\mu^2}+\ln\frac{m^2_2}{\mu^2}\right) \int 
\frac{d^4q}{\qbar^6}
\right\}. 
\eqa
The first of Eqs.~(\ref{eq:scalext}) is computed indirectly from the FDR expansion 
\bqa
\frac{q^{\alpha}q^{\beta}}{(\qbar^2-m^2)} = q^{\alpha}q^{\beta} \left\{
    \left[\frac{1}{\qbar^2}\right]
+m^2\left[\frac{1}{\qbar^4}\right]
+m^4\left[\frac{1}{\qbar^6}\right]
+m^6\left[\frac{1}{\qbar^6 (\qbar^2-m^2)}\right]
\right\},\nl
\eqa
while Eqs.~(\ref{eq:appa3}) and~(\ref{eq:appa4}) give the other three equalities.

\subsection{The result}
By summing all Feynman diagrams and performing the tensor reduction we end up with 
\bqa
\label{eq:f0}
	\F^{(2)}_0\!\!&=&
	-2 \iA{2}{2} + 4 \iA{3}{} 
	-4m^2\iA{3}{2} 
	+12m^2 \iA{4}{} 
	\nl &&
	+ 4 \iB{2}{} 
	+12m^2 \Big( 2\iB{3}{} 
	+  \iB{2}{2} \Big) 
	\nl &&
    +24m^4\Big(\iB{4}{}+\iB{3}{2}\Big) 
	+ 4 \iA{3}{2}(\mu^2|_1) 
 	\nl &&
	+ 8 \iB{3}{}(\mu^2|_1) + 4  \iB{2}{2}(\mu^2|_1)
      -2  \iB{2}{2}(\mu^2|_{12})
    \nl &&
    + 8m^2 \Big( \iB{3}{2}(\mu^2|_{2})
	- \iB{3}{2}(\mu^2|_{12})\Big)\,, \nl
\eqa
and
\bqa
\label{eq:f1}
	\F^{(2)}_1 &=& 
	+\tfrac{176}{9}m^2 \iA{3}{2} 
	-\tfrac{56}{3} m^2 \iA{4}{} \nl
      && - 4m^4 \Big( 
		\tfrac{10}{9}\iA{3}{3} -\tfrac{10}{3}\iA{4}{2} 
		+\tfrac{16}{3}\iA{5}{}
	\Big) \nl
      &&+ 4m^6  \Big( 
		\tfrac{10}{3}\iA{4}{3} +\;4\,\iA{5}{2} 
		-\tfrac{20}{3}\iA{6}{}
	\Big) \nl
      &&-\tfrac{320}{9}m^2 \iB{3}{} 
	-\tfrac{136}{9}m^2  \iB{2}{2} \nl
&& -\tfrac{176}{3}m^4\Big(\iB{4}{}+\iB{3}{2}\Big) \nl
      &&-\tfrac{224}{3}m^6 \Big(\iB{5}{}+\iB{4}{2}
				+\tfrac{1}{2}\iB{3}{3}\Big) \nl
      &&-\tfrac{160}{3}m^8  \Big(\iB{6}{}+\iB{5}{2}+\;\;\iB{4}{3}\Big) \nl
      &&- 8 m^2 \iA{3}{3}(\mu^2|_1) 
	- 8 m^4 \iA{3}{4}(\mu^2|_1)  \nl 
      &&+ \tfrac{64}{9}m^2  \iB{3}{2}(\mu^2|_{2})
	+ \tfrac{80}{9}m^2  \iB{3}{2}(\mu^2|_{12}) \nl
      &&-\,16\; m^4\Big( \iB{4}{2}(\mu^2|_2)
	-   \iB{4}{2}(\mu^2|_{12}) \Big) \nl
      &&-\tfrac{64}{3}m^6 \Big( \iB{5}{2}(\mu^2|_2)
	-   \iB{5}{2}(\mu^2|_{12}) \Big) \,.
\eqa
The final result in Eq.~(\ref{eq:expected}) follows by inserting the expressions of the scalar and extra integrals computed in subsection~\ref{sec:bblo}.

A few remarks are in order. At two loops the one-to-one correspondence between DR and FDR is lost and it is no longer true that FDR integrals are obtained from DR ones after subtracting poles (and universal constants). For example, if we were to interpret the integrals appearing in Eq.~(\ref{eq:f0}) as dimensionally regulated ones, we would not get zero and a $1/\epsilon$ pole would even remain! 
 Differences already start at the level of the basic two-loop scalar integral.
The DR counterpart of Eq.~(\ref{eq:basic}) reads~\cite{vanderBij:1983bw}
\bqa 
&&\mur^{-2 \epsilon}
 \int  d^nq_1 d^nq_2 \frac{1}{(q^2_1-m^2_1)^2 (q^2_2-m^2_2) q^2_{12}}
\nl &&
= \pi^4 
\left\{-{\rm Li_2}\left(1-\frac{m^2_2}{m^2_1}\right)
-\ln^2 \frac{\mur^2}{m^2_1}   
-\ln \frac{\mur^2}{m^2_1}+ {\rm constant}   
\right\}, 
 \eqa 
with a different coefficient in front of the $\ln^2$.
This can be understood because two different mechanisms to preserve gauge invariance are used by DR and FDR, the latter avoiding an order-by-order renormalization.  Another advantage of FDR is that the {\em same} formulae for the
scalar one-loop functions can be used also when they combine to form a factorizable two-loop integral. Differently stated, Eq.~(\ref{eq:scalarint12}) is simply the product of two integrals of the kind given in Eq.~(\ref{eq:scalarint1}). This does not happen in DR, where terms of ${\cal O}(\epsilon)$ must be added to the one-loop functions appearing in a two-loop calculation.
Note also that there is no need of renormalizing $\F^{(2)}_0$ and $\F^{(2)}_1$. This directly follows from the discussion in subsection~\ref{fdrvsdr}. FDR renormalization amounts to the mere operation of fixing results in terms of physical quantities, and since the top mass disappears due to the limit $m_{\rm top} \to \infty$, no fixing is needed.
This {\em is not} the case when using DR, where renormalization is required in order to compensate spurious ${\epsilon}/{\epsilon}$ constants generated in the limit $n \to 4$.
The situation is analyzed in the next subsection.
\subsection{Renormalization}
\label{renorm}
\begin{figure}[t]
\begin{center}
	\includegraphics[width=0.43\textwidth]{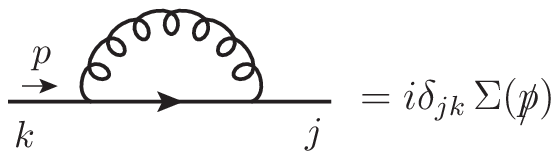}
\end{center}
\caption{\small Top self-energy at $\ord(\alpha_S)$.} \label{Top_self_energy}
\end{figure}

Here we demonstrate that if we insist with an order-by-order renormalization we obtain a vanishing contribution to $\F^{(2)}_0$ and $\F^{(2)}_1$. At $\ord(\alpha_S)$ 
the bare ($m_0$) and physical ($m$) top masses satisfy the relation 
\bqa
\label{eq:splitm}
	m_0 = m+\delta m\,,~~~\delta m = \Sigma(m)\,,
\eqa
where $\Sigma(\slashed{p})$ is the top self-energy depicted in Fig.~\ref{Top_self_energy} and
\bqa
\label{eq:delta}
\Sigma({m})= m \frac{\alpha_{S}}{3 \pi}\left(3 \ln \frac{m^2}{\mur^2} -5\right)\,.
\eqa

\begin{figure}[t]
\begin{center}
\includegraphics[width=0.85\textwidth]{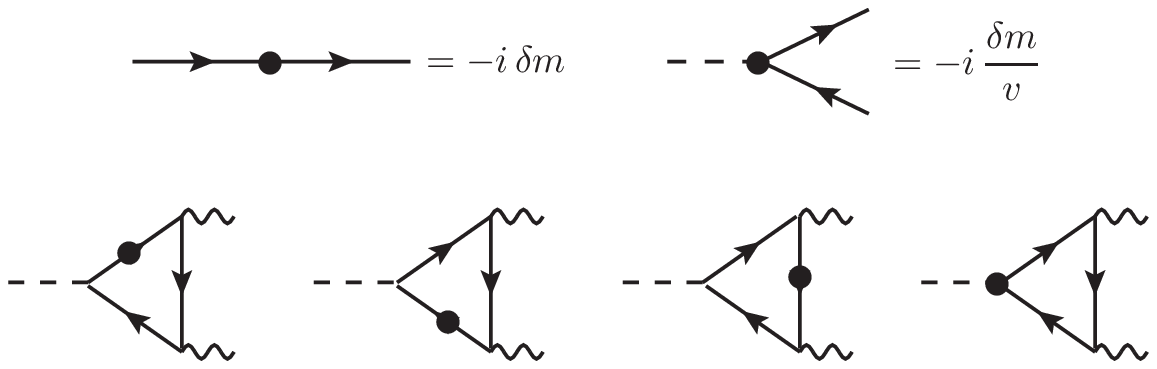}
\caption{One-loop counterterms and diagrams generated by Eq.~(\ref{eq:splitm}).}\label{CT_feyn_rules}
\end{center}
\end{figure}

This gives the one-loop counterterms and diagrams of Fig.~\ref{CT_feyn_rules},   which generate a contribution to $\F^{(2)}_0$ and $\F^{(2)}_1$ proportional to
\bqa
\F^{(2)}_{0,{ct}} = i\,\delta m\, C_{0,{ct}}\,~~~{\rm and}~~~
\F^{(2)}_{1,{ct}} = i\,\delta m\, C_{1,{ct}} \,.
\eqa

One computes 
\bqa
		C_{0,ct} &=&
		8 m^2 \iO{3} + 12 m^4 \iO{4}+ 4 \iO{3}(\mu^2) = 0\,, \nl
		C_{1,ct} &=&
		-\frac{16m^2}{3} \Big( \iO{3} + 4 m^2 \iO{4}+5 m^4 \iO{5}+5 m^6 \iO{6}  \Big) = 0\,.\nl
\eqa
Therefore renormalization does not have any effect.

It is worth mentioning that in DR 
\bqa
	C_{0,\text{ct}} \big|_{\text{DR}} =  0\,~~~{\rm and}~~~
	C_{1,\text{ct}} \big|_{\text{DR}} = \ord(\epsilon)\,,
\eqa
so that $C_{1,\text{ct}}\big|_{\text{DR}}$ contributes to the amplitude when multiplied against the $1/\epsilon$ pole contained in $\delta m\big|_{\text{DR}}$ (the DR counterpart of $\delta m$\,\footnote{$\delta m\big|_{\text{DR}}$ in Dimensional Reduction is obtained from Eq.~(\ref{eq:delta}) through the replacement 
$\ln \mur^2 \to \ln \mur^2 + \Delta$, with $\Delta$ given in 
footnote~\ref{foot1}.}), and renormalization is necessary.

\section{Conclusions}
We have presented the first two-loop calculation ever performed in FDR.  The
${\cal O}(\alpha_S)$ corrections to the $H \to \gamma \gamma$ amplitude --\,mediated by an infinitely heavy top loop\,-- have been computed in a fully four-dimensional fashion. This example has allowed us to show that FDR is an approach to loop calculations in which
\begin{itemize} 
\item gauge invariance is preserved;
\item order-by-order renormalization is avoided;
\item a finite renormalization is only needed to fix the parameters of the theory in terms of experimental observables;
\item $\ell$-loop integrals are directly re-usable in ($\ell$+1)-loop calculations, with no need of further expanding in $\epsilon$.
\end{itemize} 
In addition, we have described how infrared divergences can be dealt with within the same four-dimensional framework used to cope with the ultraviolet infinities.

We have also demonstrated that DR and FDR are not related in a direct way --\,beyond one loop\,-- since FDR integrals cannot be interpreted any longer as DR ones devoid of the pole part.
Due to its four-dimensionality we envisage a great potential of FDR in further simplifying NNLO computations. More investigation is needed in this direction, that we plan to undertake in the near future. 

\acknowledgments
This work was supported by the European Commission through contracts ERC-2011-AdG No 291377 (LHCtheory) and PITN-GA-2012-316704 (HIGGSTOOLS). We also
thank the support of the MICINN project FPA2011-22398 (LHC@NLO) and the Junta de
Andaluc\'ia project P10-FQM-6552.

\appendix
\section{FDR and shift invariance}
\label{appshift}
In this appendix we demonstrate that, for positive integers $\alpha$, $\beta$, $\gamma$ and $\delta$, 
\bqa
\label{oneloopsh}
&&\int[d^4q]\frac{1}{( q^2   -m^2 -\mu^2)^\alpha} =  
\int[d^4q] \frac{1}{((q+p)^2-m^2 -\mu^2)^\alpha}
\eqa
and
\bqa
\label{twoloopsh}
&&\qquad\qquad
\int
\frac{[d^4q_1][d^4q_2] }{(q_1^2   -m_1^2    -\mu^2)^\beta 
        ( q_2^2   -m_2^2    -\mu^2)^\gamma 
        ( q_{12}^2 -m_{12}^2 -\mu^2)^\delta} = \nl
&&
\int
\frac{[d^4q_1][d^4q_2]}{((q_1+p_1)^2   -m_1^2    -\mu^2)^\beta 
        ( (q_2+p_2)^2   -m_2^2    -\mu^2)^\gamma 
        ( (q_{12}+p_{12})^2 -m_{12}^2 -\mu^2)^\delta}\,, \nl 
\eqa
where $q_{12}= q_1+q_2$ and $p_{12}= p_1+p_2$.
Since integrals of polynomials in the integration variables vanish, the divergent parts of any one- or two-loop FDR integral can be written --\,after expanding in the external momenta\,--  in terms of the four cases
\bqa
\alpha= 1\,,~~
\alpha= 2\,,~~
\beta= \gamma= \delta = 1\,,~~{\rm and}~~\beta=\gamma= 1~{\rm with}~\delta > 1\,.
\eqa
In all the other cases Eqs.~(\ref{oneloopsh}) and~(\ref{twoloopsh}) coincide with finite integrals, for which shift invariance trivially holds.

We start proving Eq.~(\ref{oneloopsh}) with $\alpha = 1$.
By using the shorthand notation
\bqa
\dbar = (q^2-m^2-\mu^2)\,,~~\sbar = (q+p)^2-m^2-\mu^2\,,
\eqa
one writes the FDR expansions of the two sides of the equation as 
\bqa
\frac{1}{\dbar} &=&
                  \left[\frac{1}{\qbar^2} \right]
+                 \left[\frac{m^2}{\qbar^4}\right]
+                 J_{\rm F,1}(q)\,, \nl
\frac{1}{\sbar} &=& 
                  \left[\frac{1}{\qbar^2} \right]
+                 \left[\frac{m^2-p^2-2(q \cdot p)}{\qbar^4}\right]
+                4\left[\frac{(q \cdot p)^2}{\qbar^6}\right]
+                 J^{\prime}_{\rm F,1}(q)\,. 
\eqa
Then
\bqa
\label{eqpro2}
\int[d^4q] \frac{1}{{\dbar}} &=& 
\lim_{\mu \to 0} \mur^{-\epsilon}
\left(
\int d^nq  \frac{1}{\dbar}-
\int d^nq  \frac{m^2}{\qbar^4}   -
\int d^nq  \frac{1}{\qbar^2} 
\right) \nl 
&=&
\lim_{\mu \to 0} \mur^{-\epsilon}
\left(
\int d^nq  \frac{1}{\sbar}-
\int d^nq  \frac{m^2}{\qbar^4}  -
\int d^nq  \frac{1}{(q+p)^2-\mu^2}
\right)\,,
\eqa
where the shift invariance of the dimensionally regulated integrals over
$1/\dbar$ and $1/\qbar^2$ has been used. 
By now expanding the last integrand one obtains
\bqa
\label{eq:quadr}
\frac{1}{(q+p)^2-\mu^2}-
\frac{1}{\qbar^2}=
-\frac{p^2+2(q \cdot p)}{\qbar^4}
+ 4 \frac{(q \cdot p)^2}{\qbar^6} + {\cal O}(p^3)\,.
\eqa
Since the l.h.s. of Eq.~(\ref{eq:quadr}) vanishes upon integration at any order in $p$, the same happens for the combination 
\bqa
-\frac{p^2+2(q \cdot p)}{\qbar^4}
+ 4 \frac{(q \cdot p)^2}{\qbar^6}\,.
\eqa
The last integral in Eq.~(\ref{eqpro2}) can then be rewritten as
\bqa
\mur^{-\epsilon} \int d^nq  \frac{1}{(q+p)^2-\mu^2} =
\mur^{-\epsilon} \int d^nq \left(
\frac{1}{\qbar^2}
-\frac{p^2+2(q \cdot p)}{\qbar^4}
+ 4 \frac{(q \cdot p)^2}{\qbar^6}\right)\,,
\eqa
so that
\bqa
\label{eqeqe}
\int[d^4q] \frac{1}{{\dbar}} = \lim_{\mu \to 0} 
\int d^4q  J^{\prime}_{\rm F,1}(q) = \int[d^4q] \frac{1}{{\sbar}}\,,
\eqa
which proves Eq.~(\ref{oneloopsh}) with $\alpha = 1$.
The case $\alpha = 2$ is proven by taking the derivative of Eq.~(\ref{eqeqe}) with respect to $m^2$. 

We now deal with the case $\beta= \gamma= \delta= 1$.
The FDR expansion of the l.h.s. of Eq.~(\ref{twoloopsh}) is given by 
Eq.~(\ref{eq:ex2}). As for the r.h.s., we introduce
\bqa
\sbar_i   = (q_i+p_i)^2-m_i^2-\mu^2\,~~~{\rm and}~~~N_i= m^2_i-p^2_i-2 (q_i \cdot p_i )\,,
\eqa
in terms of which the expansion reads
\bqa
\label{eqrec2}
\frac{1}{\sbar_1 \sbar_2 \sbar_{12}} &=&
 \left[\frac{1}{\qbar^2_1 \qbar^2_2 \qbar^2_{12}} \right] 
+  \left[\frac{N_1}{\qbar^4_1 \qbar^2_2 \qbar^2_{12}} \right]
+  \left[\frac{N_2}{\qbar^2_1 \qbar^4_2 \qbar^2_{12}} \right]
+\left[\frac{N_{12}}{\qbar^2_1 \qbar^2_2 \qbar^4_{12}} \right] 
+ 4      \left[\frac{(q_1 \cdot p_1)^2}{\qbar^6_1 \qbar^2_2 \qbar^2_{12}} \right] 
\nl &&
+ 4      \left[\frac{(q_2 \cdot p_2)^2}{\qbar^2_1 \qbar^6_2 \qbar^2_{12}} \right] 
+ 4 \left[\frac{(q_{12} \cdot p_{12})^2}{\qbar^2_1 \qbar^2_2 \qbar^6_{12}} \right] 
+ 4\left[\frac{(q_1 \cdot p_1)(q_2 \cdot p_2)}{\qbar^4_1 \qbar^4_2 \qbar^2_{12}} \right] 
\nl &&
+ 4\left[\frac{(q_1 \cdot p_1)(q_{12} \cdot p_{12})}{\qbar^4_1 \qbar^2_2 \qbar^4_{12}} \right] 
+ 4\left[\frac{(q_2 \cdot p_2)(q_{12} \cdot p_{12})}{\qbar^2_1 \qbar^4_2 \qbar^4_{12}} \right]
\nl &&
+ \left( 
\frac{N^2_1}{\qbar^4_1 \sbar_1}-4\frac{(q_1 \cdot p_1)^2}{\qbar^6_1}
\right)\left[\frac{1}{\qbar^4_2}\right] 
+ \left( 
\frac{N^2_2}{\qbar^4_2 \sbar_2}-4\frac{(q_2 \cdot p_2)^2}{\qbar^6_2}
\right)\left[\frac{1}{\qbar^4_1}\right]
\nl &&
+ \left( 
\frac{N^2_{12}}{\qbar^4_{12} \sbar_{12}}-4\frac{(q_{12} \cdot p_{12})^2}{\qbar^6_{12}}
\right)\left[\frac{1}{\qbar^4_1}\right] + J^\prime_{{\rm F},2}(q_1,q_2) \,. 
\eqa
Then, by rewriting
\bqa
\frac{m_i^4}{\bar D_i\bar{q}_i^4}= \frac{1}{\dbar_i}-\frac{1}{\qbar_i^2}-\frac{m^2_i}{\qbar_i^4}
\eqa
and shifting all the $\dbar_i$ and the quadratically divergent integrals, Eq.~(\ref{eq:ex2}) produces
\bqa
\int[d^4q_1][d^4q_2] \frac{1}{\dbar_1 \dbar_2 \dbar_{12}} &=&
\lim_{\mu \to 0} \mur^{-2 \epsilon}
\int d^nq_1 d^nq_2  
\left(
\frac{1}{\sbar_1 \sbar_2 \sbar_{12}} \right.
\nl &&
-\frac{1}{((q_1+p_1)^2-\mu^2)((q_2+p_2)^2-\mu^2)((q_{12}+p_{12})^2-\mu^2)}
\nl &&
- m_1^2
{ \left[
\frac{1}{\bar{q}_1^4\bar{q}_2^2 \bar{q}_{12}^2}
 \right]}
- m_2^2
{ \left[
\frac{1}{\bar{q}_1^2 \bar{q}_2^4\bar{q}_{12}^2}
 \right]}
- m_{12}^2
{ \left[
\frac{1}{\bar{q}_1^2 \bar{q}_2^2\bar{q}_{12}^4}
 \right]}
\nl &&
-\left(\frac{1}{\sbar_1}-\frac{1}{(q_1+p_1)^2-\mu^2}-\frac{m^2_1}{\qbar_1^4}\right)
{ \left[\frac{1}{\bar{q}_2^4} \right]}
\nl &&
- \left(\frac{1}{\sbar_2}-\frac{1}{(q_2+p_2)^2-\mu^2}-\frac{m^2_2}{\qbar_2^4}\right)
{ \left[\frac{1}{\bar{q}_1^4} \right]}
\nl &&
-\left(\frac{1}{\sbar_{12}}-\frac{1}{(q_{12}+p_{12})^2-\mu^2}-\frac{m^2_{12}}{\qbar_{12}^4}\right)
\left.{ \left[\frac{1}{\bar{q}_1^4} \right]} \right)\,.
\eqa
An expansion up to ${\cal O}(p^2_1)$, ${\cal O}(p^2_2)$ and ${\cal O}(p_1 p_2)$ of the second line and of the terms $$1/((q_i+p_i)^2-\mu^2)$$ in the last three lines produces extra integrands which --\,by the same argument used at one-loop\,-- vanish upon integration. The addition of such terms reconstructs $J^\prime_{{\rm F},2}(q_1,q_2)$ as given in Eq.~(\ref{eqrec2}), so that
\bqa
\int[d^4q_1] [d^4q_2] \frac{1}{\dbar_1 \dbar_2 \dbar_{12}} = \lim_{\mu \to 0} 
\int d^4q_1 d^4q_2  J^\prime_{\rm F,2}(q_1,q_2) = \int[d^4q_1] [d^4q_2] \frac{1}{{\sbar_1 \sbar_2 \sbar_{12}}}\,.
\eqa
Finally, deriving with respect to $m^2_{12}$ demonstrates the last case.

With more loops the proof follows the same reasoning: the mismatch between the FDR expansion of shifted and unshifted integrands is cured by vanishing integrals obtained by expanding the polynomially divergent integrals in $J_{\rm INF}(q_1,\ldots, q_\ell)$ at the relevant order in $p$, as in Eq.~(\ref{eq:quadr}).

\section{Feynman rules}
\label{app0}
\begin{figure}[h]
\begin{center}
\includegraphics[width=0.8\textwidth]{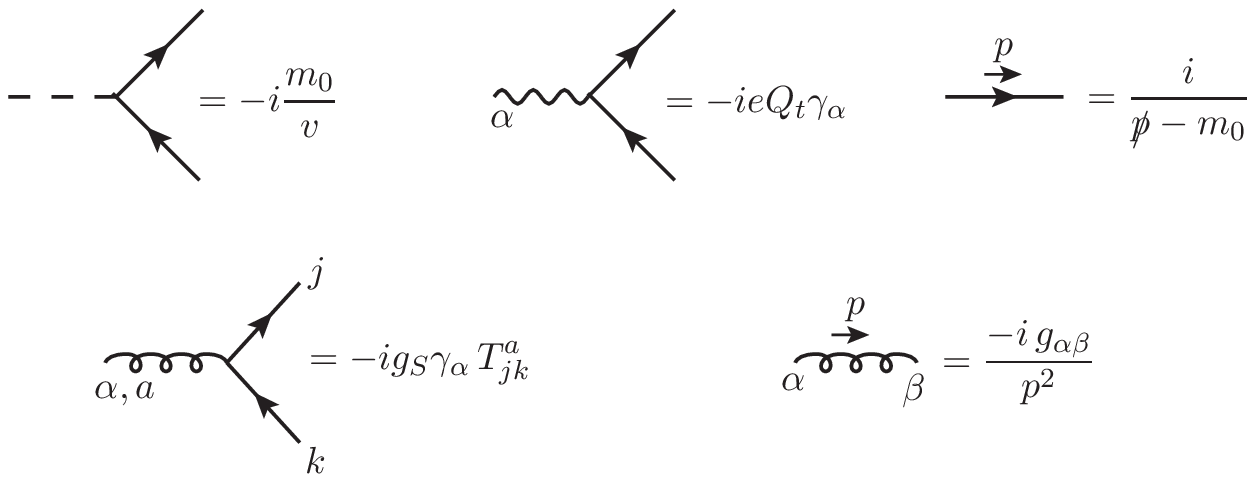}
\caption{Feynman rules used in the computation of $H \to \gamma \gamma$ at $\ord(\alpha_S)$.} \label{EW_Feynman_rules}
\end{center}
\end{figure}

For completeness we list, in Fig.~\ref{EW_Feynman_rules}, the Feynman rules used in the calculation. $Q_t$, $m_0$  and $v$ are the top quark charge, the top bare mass and the vacuum expectation value of the Higgs field, respectively.

\section{A few FDR defining expansions}
\label{appa}
We collect here the two-loop FDR defining expansions used throughout the paper. Denominators are defined in Eq.~(\ref{eq:2lden}) and divergent integrands are written in square brackets.
\begin{enumerate}
\item Expansion for $\displaystyle \int [d^4q_1][d^4q_2] \frac{q^\alpha_1 
q^\beta_1}{\dbar^3_1 \dbar_2 \dbar_{12}}$:

\bqa
\label{eq:appa1}
\frac{q^\alpha_1 q^\beta_1}{\dbar^3_1 \dbar_2 \dbar_{12}} &=&
q^\alpha_1 q^\beta_1 
\left\{
\left[\frac{1}{\qbar^6_1 \qbar^2_2 \qbar^2_{12}}  \right]
  +\left(
        \frac{1}{\bar D^3_1}
       -\frac{1}{\bar{q}_1^6}
    \right)
\left(
 \left[\frac{1}{\qbar_2^4}\right]
 -\frac{q_1^2+2(q_1 \cdot q_2)}{\bar{q}_2^4 \bar{q}_{12}^2} 
\right)
\right. \nl
 &&+\left.\frac{1}{\bar D_{1}^3 \bar{q}_2^2\bar D_{12} }
 \left(
 \frac{m_2^2}{\bar D_{2}}+\frac{m_{12}^2}{\bar{q}_{12}^2}
 \right) 
\right \}\,.
\eqa

\item Expansion for $\displaystyle \int [d^4q_1][d^4q_2] \frac{1}{\dbar^2_1 \dbar_2 \qbar^2_{12}}$:

\bqa
\label{eq:basintexp}
\frac{1}{\dbar^2_1 \dbar_2 \qbar^2_{12}} &=& 
\left[\frac{1}{\qbar^4_1 \qbar^2_2 \qbar^2_{12}}  \right]
+   
\left(
 \frac{m^2_1}{\dbar_1 \qbar_1^4}+\frac{m^2_1}{\dbar^2_1 \qbar_1^2}
\right)
\left(
\left[
\frac{1}{\qbar_2^4}
\right]
-\frac{q_1^2+2(q_1 \cdot q_2)}{\qbar_2^4 \qbar_{12}^2}
\right) \nl
&&+ \frac{m_2^2}{\dbar^2_1 (\dbar_2\qbar^2_2) \qbar^2_{12}}\,.
\eqa

\item Expansion for $\displaystyle \int [d^4q_1][d^4q_2] \frac{q^\alpha_1 
q^\beta_1}{\dbar^2_1 \dbar^2_2 \qbar^2_{12}}$:

\bqa
\label{eq:appa3}
\frac{q^\alpha_1 q^\beta_1}{\dbar^2_1 \dbar^2_2 \qbar^2_{12}} &=&
q^\alpha_1 q^\beta_1 
\left\{
\left[\frac{1}{\qbar^4_1 \qbar^4_2 \qbar^2_{12}}  \right]
+ 
\left(
 \frac{m^2_2}{\dbar_2 \qbar_2^4}+\frac{m^2_2}{\dbar^2_2 \qbar_2^2}
\right)
\left(
   \left[\frac{1}{\qbar_1^6}\right]
  -\frac{q_2^2+2(q_1 \cdot q_2)}{\bar{q}_1^6 \bar{q}_{12}^2} 
\right)
\right. \nl
 &&+\left.
   \left(
         \frac{1}{\bar D^2_1}
       - \frac{1}{\bar{q}_1^4}
    \right)
\frac{1}{\bar D_{2}^2 \qbar_{12}^2 }
\right \}\,.
\eqa

\item Expansion for $\displaystyle \int [d^4q_1][d^4q_2] \frac{q^\alpha_{12} 
q^\beta_{12}}{\dbar^2_1 \dbar^2_2 \qbar^2_{12}}$:

\bqa
\label{eq:appa4}
\frac{q^\alpha_{12} q^\beta_{12}}{\dbar^2_1 \dbar^2_2 \qbar^2_{12}} &=&
q^\alpha_{12} q^\beta_{12} 
\left\{
\left[\frac{1}{\qbar^4_1 \qbar^4_2 \qbar^2_{12}}  \right]
 +\left[\frac{1}{\qbar_{12}^6}\right]
\left(
\left(
 \frac{m^2_1}{\dbar_1 \qbar_1^4}+\frac{m^2_1}{\dbar^2_1 \qbar_1^2}
\right)
+
\left(
 \frac{m^2_2}{\dbar_2 \qbar_2^4}+\frac{m^2_2}{\dbar^2_2 \qbar_2^2}
\right)
\right)
\right. \nl
&& \left. + \frac{1}{\qbar^2_{12}}
\left(
  \left(
         \frac{1}{\bar D^2_1}
       - \frac{1}{\bar{q}_1^4}
  \right) 
  \left(
         \frac{1}{\bar D^2_2}
       - \frac{1}{\bar{q}_2^4}
  \right)
+
  \left(
         \frac{1}{\qbar^4_1}
       - \frac{1}{\qbar^4_{12}}
  \right)
  \left(
         \frac{1}{\bar D^2_2}
       - \frac{1}{\bar{q}_2^4}
  \right) \right. \right. \nl
&& \left. \left. +
  \left(
         \frac{1}{\qbar^4_2}
       - \frac{1}{\qbar^4_{12}}
  \right)
  \left(
         \frac{1}{\bar D^2_1}
       - \frac{1}{\bar{q}_1^4}
  \right)
\right)
\right \}\,.
\eqa

\end{enumerate}

\section{The basic two-loop scalar integral}
\label{appb}
In this appendix, we demonstrate that the basic two-loop scalar integral in
Eq.~(\ref{eq:eqfunint}) reads
\bqa
\label{eq:basic}
\big[ \,2 m_1\,|\, m_2\,|\, 0 \,\big] &\equiv& \int  [d^4q_1][d^4q_2] \frac{1}{\dbar^2_1 \dbar_2 \qbar^2_{12}}
\nl
&=& \pi^4 
\left\{f-{\rm Li_2}\left(1-\frac{m^2_2}{m^2_1}\right)
-\frac{1}{2}\ln^2 \frac{\mur^2}{m^2_1}   
-\ln \frac{\mur^2}{m^2_1}   
\right\} \,, 
\eqa
with
\bqa
\label{eq:f}
f= \frac{i}{\sqrt{3}} 
\left(
 {\rm Li_2}(e^{i\frac{\pi}{3}})
-{\rm Li_2}(e^{-i\frac{\pi}{3}})
\right)\,.
\eqa
A direct integration of the finite part of its FDR defining expansion
--\,Eq.~(\ref{eq:basintexp})\,-- gives
\bqa
\big[ \,2 m_1\,|\, m_2\,|\, 0 \,\big] = m^2_2 I_2(m_1,m_2)-m^2_1I_1(m_1)\,,
\eqa
where
\bqa
I_2(m_1,m_2) &=&  \lim_{\mu \to 0} \int d^4q_1 d^4q_2 
\frac{1}{\dbar^2_1 (\dbar_2\qbar^2_2) \qbar^2_{12}}~~~{\rm and} \nl
I_1(m_1)     &=&  \lim_{\mu \to 0} \int d^4q_1 d^4q_2 
\frac{q_1^2+2(q_1 \cdot q_2)}{\qbar_2^4 \qbar_{12}^2}
\left(
 \frac{1}{\dbar_1 \qbar_1^4}+\frac{1}{\dbar^2_1 \qbar_1^2}
\right) \,.
\eqa
By power counting --\,due to the presence of ${1}/{q^4_i}$ terms\,-- a logarithmic dependence on $\mu$ is expected in $I_1(m_1)$,  while $\mu$ can be immediately set to zero in $I_2(m_1,m_2)$.
A natural split is then obtained in FDR:  $I_2(m_1,m_2)$ only depends on 
\bqa
r_{12}= \frac{m_1^2}{m_2^2}\,,
\eqa
and $I_1(m_1)$ on
\bqa
\rho_1= \frac{\mu^2}{m^2_1}\,,
\eqa
so no difficult integral containing both ratios needs to be computed.
A simple Feynman parametrization produces
\bqa
\label{}
I_2(m_1,m_2)&=&  \int_0^1 dz \int d^4q_1 d^4q_2
\frac{1}{D^2_1 (q^2_2-m^2_2 z)^2 q^2_{12}} \nl
 &=&  
\frac{\pi^4}{m^2_2}  
\int_0^1 dz
\int_0^1 dx
\int_0^1 dy 
\frac{y}{xyz+r_{12}(1-y)}
=
\frac{\pi^4}{m^2_2} 
\left\{ 
\frac{\pi^2}{6}
-{\rm Li_2}\left( \frac{r_{12}-1}{r_{12}}\right)
\right\}\,, \nl
\eqa
and
\bqa
I_1(m_1)&=& 2 \lim_{\mu \to 0} \int_0^1 dz \int d^4q_1 d^4q_2 
\frac{q_1^2+2(q_1 \cdot q_2)}{(\qbar^2_1-m^2_1 z)^3\qbar_2^4 \qbar_{12}^2} \nl
&=& \frac{2 \pi^4}{m_1^2}
\lim_{\mu \to 0} 
\int_0^1 dz
\int_0^1 dx
\int_0^1 dy 
\frac{(2x-1)y^2x}{(z+\rho_1)x(1-x)y+\rho_1(1-y)} \nl
&=& \frac{\pi^4}{m_1^2}
\left\{
\frac{\pi^2}{6}-f
+\frac{1}{2}\ln^2\rho_1 
+\ln \rho_1   
\right\}\,,
\eqa
from which Eq.~(\ref{eq:basic}) follows.
The same result can be derived as a finite combination of divergent
integrals. From Eq.~(\ref{eq:basintexp}), by using DR,
\bqa
\big[ \,2 m_1\,|\, m_2\,|\, 0 \,\big] &=& 
\mur^{-2 \epsilon}
\int d^nq_1 d^nq_2 
\frac{1}{D^2_1 D_2 q^2_{12}}
-\lim_{\mu \to 0} \mur^{-2 \epsilon}\int d^nq_1 d^nq_2 
 \left[\frac{1}{\qbar^4_1 \qbar^2_2 \qbar^2_{12}}\right] \nl
&&-\lim_{\mu \to 0}  \mur^{-2 \epsilon} \int d^nq_2 \left[ \frac{1}{\qbar^4_2}\right] 
 \int d^nq_1
\left(
\frac{m^2_1}{\dbar_1 \qbar_1^4}+\frac{m^2_1}{\dbar^2_1 \qbar_1^2}
\right)\,.
\eqa
The relevant DR integrals can be found in the appendix of~\cite{vanderBij:1983bw}. 

\bibliographystyle{spphys}       
\bibliography{paper}   
\end{document}

\end{document}